\documentclass[twocolumn]{aastex631}

\shorttitle{The host of the short GRB 080905A}
\shortauthors{Nicuesa Guelbenzu et al.}

\usepackage{xcolor}
\usepackage{enumerate}

\newcommand{\HST}{\rm{HST}}

\newcommand{\swift}{\textit{Swift}}

\def\OIII{[O\,{\sc iii}]~$\lambda$}
\def\SII{[S\,{\sc ii}]~$\lambda$}

\def\NII{[N\,{\sc ii}]~$\lambda$}

\def\HII{[H\,{\sc ii}]}
\def\Ha{H$\alpha$}
\def\Hb{H$\beta$}
\def\msun{~$M_\odot$\ }
\def\rsun{~$R_\odot$}
\newcommand{\msunyr}{$M_\odot$\,yr$^{-1}$}

\usepackage{hyperref}
\hypersetup{
bookmarks=true,     
unicode=true,       
colorlinks=true,    
linkcolor=red,      
citecolor=blue,     
filecolor=magenta,  
urlcolor=black      
}

\usepackage{colortbl,xcolor,framed}

\begin{document}


\title{VLT/MUSE and ATCA observations of the host galaxy of the \\
  short GRB 080905A at $z$=0.122}

\author[0000-0002-6856-9813]{A.~M.~Nicuesa Guelbenzu}
\author[0000-0001-8413-7917]{S.~Klose}
\affiliation{Th\"uringer Landessternwarte Tautenburg, Sternwarte 5, 07778 Tautenburg, Germany}

\author[0000-0002-1214-770X]{P. Schady}
\affiliation{Department of Physics, University of Bath, Claverton Down, Bath BA2 7AY, UK}

\author[0000-0002-1658-7681]{K. Belczynski}
\affiliation{Nicolaus Copernicus Astronomical Center, Polish Academy of Sciences, ul. Bartycka 18, 00-716 Warsaw, Poland}

\author[0000-0002-8028-0991]{D.~H.~Hartmann}
\affiliation{Department of Physics and Astronomy, Clemson University, Clemson, SC 29634, USA}

\author[0000-0001-9162-2371]{L.~K.~Hunt}
\affiliation{INAF -- Osservatorio Astrofisico di Arcetri, I-50125 Firenze, Italy}

\author[0000-0001-9033-4140]{M.~J.~Micha{\l}owski}
\affiliation{Astronomical Observatory Institute, Faculty of Physics, Adam Mickiewicz University, ul.~S{\l}oneczna 36, 60-286 Pozna{\'n}, Poland}


\begin{abstract} 

Short-GRB progenitors could come in various flavors, depending on the
nature of the merging compact stellar objects (including a
stellar-mass black hole or not) or depending on their ages (millions
or billions of years).
%
At a redshift of $z$=0.122, the nearly face-on spiral host of the short
GRB 080905A is one of the closest short-GRB host galaxies identified
so far. This made it a preferred target to explore spatially resolved
star-formation and to investigate the afterglow  position in the
context of its star formation structures.
%
We used VLT/MUSE integral-field unit observations, supplemented
by ATCA 5.5/9.0~GHz radio-continuum measurements and publicly
available HST data, to study the star-formation activity in the GRB
080905A host galaxy.
%
The MUSE observations reveal that the entire host is characterized by
strong line emission. Using the H$\alpha$ line flux, we measure for
the entire galaxy an SFR of about 1.6~\msunyr, consistent with
its non-detection by ATCA. Several individual star-forming regions
are scattered across the host. The most luminous region has a \Ha\
luminosity that is nearly four times as high as the luminosity of the
Tarantula nebula in the Large Magellanic Cloud.
%
Even though star-forming activity can be traced as close to about
3~kpc (in projection) distance to the GRB explosion site, stellar
population synthesis calculations show that none of
the \Ha-bright star-forming regions is a likely birthplace of the
short-GRB progenitor.

\end{abstract}


\section{Introduction \label{Intro} }

To better understand the diversity in gamma-ray bursts (GRBs), host-galaxy
studies  offer a powerful observational tool. Early work in this
respect goes back to times when GRB afterglows were not yet
known.  Before  1997, observations focused on a search for 'unusual' or
statistically rare galaxies in the smallest available GRB error boxes
(Interplanetary Network error boxes;
e.g., \citealt{Atteia1987ApJS...64..305A,
Hurley1993AIPC..280..769H,Hurley1994AIPC..307..653H}) in order to find
clues to the distance scale and underlying nature of the bursts
(e.g., \citealt{Boer1991A&A...249..118B, Vrba1995ApJ, Klose1996ApJ,
Larson1996ApJ...460L..95L, Hurley1997ApJ...479L.113H,
Larson1997ApJ...491...93L, Schaefer1998ApJS..118..353S}).  Although
these observations did not identify valid host candidates, they did
support the notion that bursts of extragalactic origins would be
associated  with normal host galaxies, i.e., not galaxies
characterized by unusual physical parameters.

Once the first optical afterglow of a {\it long} GRB was discovered
(\citealt{Groot1997IAUC.6584....1G, Paradijs1997Natur.386..686V}),
studies of GRB host galaxies quickly developed into a powerful
approach to reveal additional information about the nature of GRB
progenitors (e.g., \citealt{Bloom1998ApJ...507L..25B,
Djorgovski1998ApJ...508L..17D, Sokolov1999A&A...344...43S,
Fruchter2006Natur.441..463F, Thoene2014MNRAS.441.2034T,
Lyman2017MNRAS.467.1795L}).

A milestone in the exploration of GRB host galaxies was
the first identified long-GRB progenitor, GRB 980425/SN~1998bw
(\citealt{Galama1998}). Due to its very low redshift ($z$=0.0085; 
\citealt{Tinney1998IAUC}), the host became the focus of several
comprehensive observing campaigns, which provided substantial insight
into the GRB explosion site and hence the nature of the GRB progenitor
(e.g., \citealt{Hammer2006A&A...454..103H,
Christensen2008A&A...490...45C,Michalowski2009,
LeFloch2012ApJ...746....7L,
Michalowski2014A&A...562A..70M,
Michalowski2015A&A...582A..78M,
Michalowski2016A&A...595A..72M,
Arabsalmani2015MNRAS.454L..51A,
Kruehler2017}).

Long before the first afterglow of a {\it short} GRB was discovered
and localized at an arcsecond scale (GRB 050509B;
\citealt{Gehrels2005Natur}), convincing arguments were already in place that
short GRBs could be linked to merging compact stellar binaries
(for a review, see \citealt{Nakar2007PhR...442..166N}),
which naturally includes elliptical galaxies as their potential
hosts. The discovery that the host of GRB 050509B
was a giant elliptical ($z$=0.225;
\citealt{Gehrels2005Natur,Hjorth2005,Bloom2006ApJ638})
confirmed this notion. Finally, the study of the hosts of short bursts
received a substantial boost by the discovery of
a physical link between the gravitational wave event GW170817 and 
the short GRB 170817A and the identification of
its early-type host galaxy \citep[e.g.,][]{Abbott2017ApJ...848L..12A,
Coulter2017Sci...358.1556C,Hjorth2017ApJ...848L..31H,
Kasen2017Natur.551...80K,Smartt2017Natur.551...75S,Kim2017ApJ...850L..21K}.

Several years ago our group started a comprehensive observing campaign
designed to study the host-galaxy population of short GRBs in order to
characterize  their star formation activity, with particular emphasis
on deep radio-continuum observations (GRB
071227: \citealt{Nic2014ApJ}; GRB 100628: \citealt{Nicuesa2015}; GRB
050709: \citealt{Nicuesa2020}).  In \cite{Klose2019}, we provided a
first summary of this radio survey program and reported on 16 targeted
short-GRB hosts.

So far, in two cases we have been able to supplement our ATCA and VLA
radio observations by using  the Multi-Unit Spectroscopic Explorer
(MUSE; \citealt{Bacon2010SPIE.7735E..08B})
mounted at the Very Large Telescope (VLT). First
results were reported in \cite{Nicuesa2020}, where we studied in
detail the host of GRB 050709.  Here, we continue these studies and
focus on the host of the short GRB 080905A.
At a redshift of $z=0.1218\pm0.0003$ \citep{Rowlinson2010a}, the host
of GRB 080905A is among the nearest short-GRB hosts detected to
date (\citealt{Berger2014ARAA})\footnote{See also Jochen Greiner's website page at
\url{https://www.mpe.mpg.de/~jcg/grbgen.html}},
qualifying it as
one of the presently best targets for spatially resolved studies. 

The structure of this work closely follows our previous study of GRB
050709, though our conclusions about the nature of these two hosts are
very different.  Throughout this paper, we adopt a flat universe with
H$_{\rm 0}$ = 68~km s$^{-1}$ Mpc$^{-1}$, $\Omega_{\rm M}$=0.31, and
$\Omega_\Lambda$=0.69 (\citealt{Planck2016A&A...594A..13P}). Then, for
$z$=0.1218, the luminosity distance is $d_L=1.81\,\times\,10^{27}$~cm
(585 Mpc), the age of the universe is 12.13 Gyr, and 1~arcsec
corresponds to 2.25~kpc projected distance. 

\section{The Burst, Its Afterglow, and Its Host Galaxy}

\subsection{The Burst and Its Afterglow}

The \emph{Neil Gehrels Swift Observatory} (\citealt{Gehrels2004})  BAT
telescope (\citealt{Barthelmy2005}) and Fermi/GBM
(\citealt{Meegan2009ApJ...702..791M}) triggered on GRB 080905A
on 5 September 2008 at
11:58:54 UT (\citealt{Bissaldi8204,Pagani8180}). The BAT light curve
showed three peaks with a total duration of $T_{90}$ (15--350
keV)=$1.0\pm0.1$~s (\citealt{Cummings8187}). The lag time of the burst
between the BAT [5–50 keV] and the [50–100 keV] channel
was $4\pm17$~ms, consistent with zero (\citealt{Rowlinson2010a}),
which is characteristic for the short-burst population (e.g.,
\citealt{Gehrels2006Natur.444.1044G,Norris2006ApJ...643..266N,
Shao2017ApJ...844..126S,Lu2018ApJ...865..153L}).
  
A fading X-ray afterglow was found by \swift/XRT (for the
instrument description see
\citealt{Burrows2005SSRv}) at coordinates R.A., decl. (J2000)
=19:10:41.74, $-$18:52:48.8, with an error circle of 1\farcs6 
(radius, 90\% confidence; \citealt{Evans8203}),\footnote{XRT coordinates
have been refined to R.A., decl. (J2000) =
19:10:41.79, $-$18:52:48.4, with an error circle radius of 1\farcs7
(\citealt{Evans2009}); see
\url{https://www.swift.ac.uk/xrt_positions/index.php}.} but
no optical afterglow was detected with \swift/UVOT
(\citealt{Pagani8180}; for UVOT: \citealt{Roming2005SSRv}).  Even
though the burst occurred in a crowded stellar field, a  faint optical
afterglow candidate ($R\sim$24~mag) was finally discovered 8 hr
after the GRB trigger with the  Nordic Optical Telescope (NOT) and
with VLT/FORS2 at coordinates  R.A., decl. (J2000) = 19:10:41.73,
$-$18:52:47.3 ($\pm$0\farcs6;
  \citealt{Malesani8190}). Its fading was confirmed by
follow-up observations with FORS2 1.5 days after the burst, and a
host-galaxy candidate was identified (\citealt{deUgarte8195}). 

GROND, mounted at the MPG 2.2m telescope on ESO/La Silla
(\citealt{Greiner2008}), started observing the field of GRB 080905A
about 17.5 hr after the burst.
Due to visibility constraints, observations could be performed for
only 11 min. The combined $g^\prime
r^\prime i^\prime z^\prime$-band image as well as the combined
$JHK$-band image did not reveal the afterglow, even though the image depth
reached  $g'r'i'z'JHK$= 23.0, 22.8, 22.3, 21.9, 20.4, 19.9, and 19.6 (AB
magnitudes; \citealt{Nicuesa2012}). Published work later showed that at
the time of the GROND observations, the magnitude of the afterglow was
around $R_C=24$ (\citealt{Rowlinson2010a}),
about 1 mag below GROND's detection limit for an 11 min exposure. 

No emerging supernova component was detected following the burst, supporting 
a non-collapsar origin of this event (\citealt{Rowlinson2010a}).

\subsection{The GRB Host Galaxy \label{host}}

Using VLT data, the afterglow and its host were analyzed in 
detail by \cite{Rowlinson2010a}. The afterglow coordinates were
refined by these authors to R.A., decl. (J2000) = 19:10:41.71,
$-$18:52:47.62 ($\pm$0\farcs76). The afterglow was located in the
outer arm of a barred spiral galaxy at $z=0.1218\pm0.0003$.  The
projected offset of the burst from the  light center of the galaxy was
$\sim$18.5 kpc.

\cite{Rowlinson2010a} derived an absolute host magnitude of 
$M_V=-21$ mag and a mass in old stars (based on VLT $K_s$-band data)
of ($2\pm1)\,\times\, 10^{10}\,M_\odot$. Because the field is
crowded with Galactic foreground stars, no star-formation rate (SFR)
could be derived for this galaxy by these authors.

\begin{figure}[t!]
\includegraphics[width=0.48\textwidth]{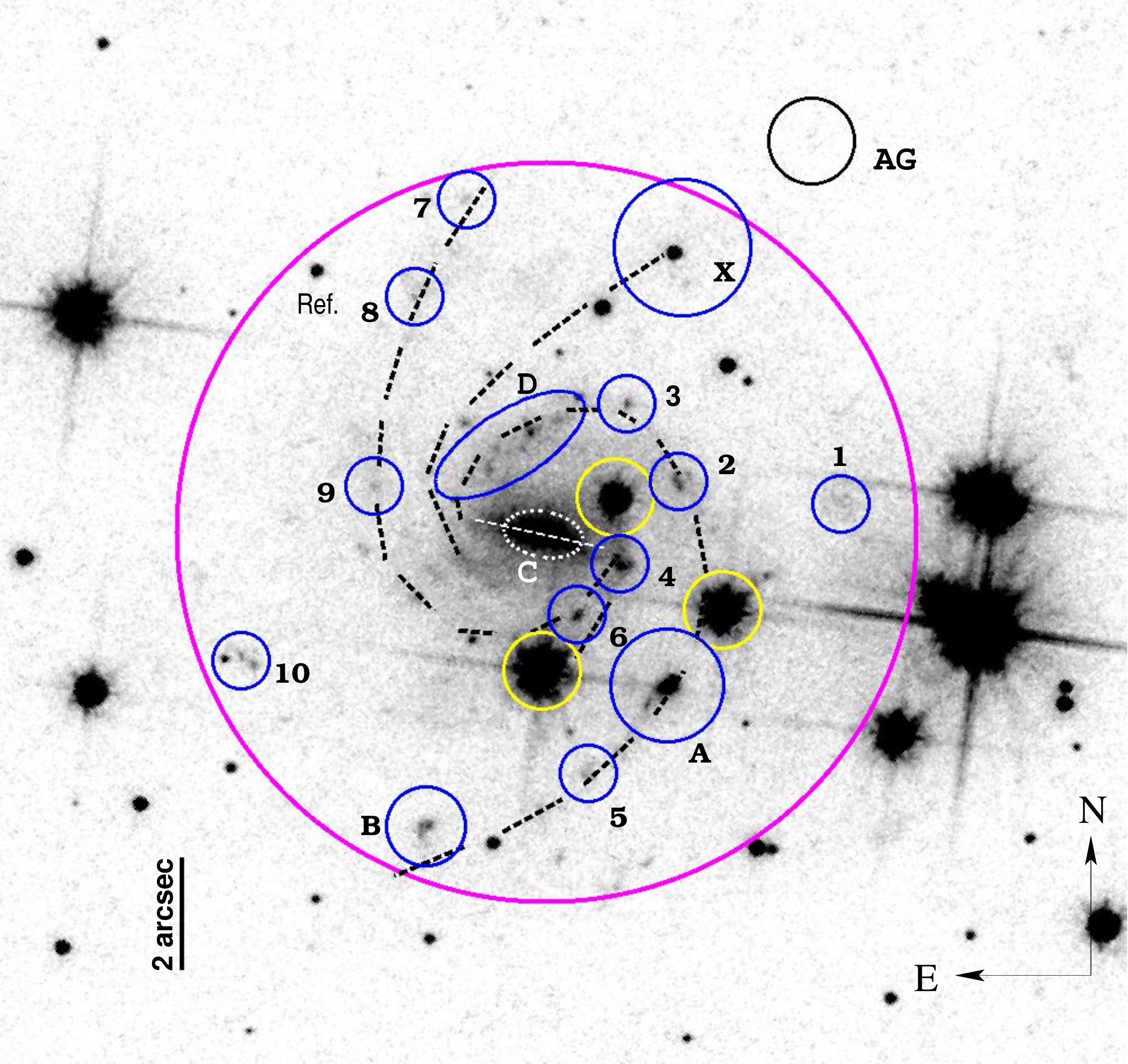}
\caption {HST/F606W image (program ID 12502; PI: A. Fruchter) of the
  field of GRB 080905A taken on 2012-04-14.  The afterglow (AG, error
  circle 0\farcs76 radius, in black; \citealt{Rowlinson2010a}) was located
  $\sim$8\farcs5 from  the center of its host. Three bright
    Galactic foreground stars are encircled by a yellow line. Shown in blue
  are several star-forming complexes that can be identified by their
  strong \Ha \ emission  (Sect.~\ref{Results}). They are numbered
  according to their R.A. (Table~\ref{Tab:A1});  
  the three \Ha-brightest regions are
  labeled with letters (A, B, D), however.
  Region C is the galaxy's bar, region X is the star-forming region
  nearest (in projection) to the AG position (Sect.~\ref{SFR}).  The large
  circle in magenta encompasses the region for which  we measure the
  total star formation rate (Sect.~\ref{SFR}). Also shown is a
  reference star (about 2 arcsec east of \#8) for which we
  measure on the HST image R.A., decl. (J2000) = 19:10:42.323,
  $-$18:52:49.88 $(\pm$0\farcs08). The spiral pattern of the galaxy is
  sketched by broken lines.}
\label{fig:HST_080905} 
\end{figure}

The field was also observed by HST on two different
occasions, in October 2011 and April 2012 (program ID 12502; PI: A.
Fruchter).  Figure~\ref{fig:HST_080905} shows the F606W image of the
field (archival file name: ibsh16030$_-$drz.fits),  where the barred spiral
morphology of the galaxy appears clearly
(\citealt{Fong2013ApJ...776...18F}).\footnote{At the given redshift,
the F606W filter includes the H$\beta$ and the \OIII 5007 emission lines.}
Its general shape strongly resembles 
NGC 1365 in the Fornax cluster (for an elegant mathematical
characterization of the spiral pattern of this galaxy
see \citealt{Ringermacher2009MNRAS.397..164R}). 
 
\subsection{Is the Suspected Host a Foreground Galaxy? \label{background} }

For GRB 080905A no redshift measurement
could be obtained via afterglow spectroscopy. This 
raises the question of the robustness of the suspected
association between GRB
080905A and the $z$=0.122 spiral galaxy, an issue that has already
been addressed by other authors. In brief, there is no ultimate proof
that the burst was associated with this galaxy. However, there is no
convincing counter argument either.
There are at least two studies published in the literature
that have to be mentioned in this respect.

(i) Deep HST/F160W imaging revealed a faint
(AB$\sim$26 mag) galaxy close to the position of the optical AG
(offset $\sim$0\farcs7; \citealt{Fong2013ApJ...776...18F}).
Its small angular size ($<$1 arcsec) might be indicative of 
a redshift $z>$1. According to \cite{Fong2013ApJ...776...18F}, the
probability of a chance coincidence $p$
between the burst and this galaxy is 0.08, compared to
$p$=0.01 for the $z=0.122$ galaxy. Even though this defines 
the faint galaxy as another host-galaxy candidate, the much
smaller $p$ favors the assumption that the large spiral is the host.

(ii) \cite{Davanzo2014MNRAS.442.2342D}
performed a statistical analysis of all short GRBs
and found that for a redshift of 0.122 the burst 080905A is an outlier
among the short-GRB ensemble 
in the $E_{\rm peak} - E_{\rm iso}$ diagram. Given its observed
$E_{\rm peak}$, the isotropic equivalent energy calculated by these
authors ($3.2 \pm 0.3\,\times\,10^{49}$ erg) lies more than a factor
of 10 below the expectations.\footnote{We note, however, that
\cite{Fong2015ApJ...815..102F}
calculated a substantially higher value for
$E_{\rm iso}$, namely $2\,\times\,10^{50}$ erg.}
As pointed out by these
authors, a higher redshift could solve this apparent conundrum.  However,
at that time, their
statistics were based on a relatively small comparison
sample (10 short
bursts). More recent data, including many more short bursts,
no longer support the conclusion that
GRB 080905A is an outlier in the $E_{\rm peak} - E_{\rm iso}$ diagram
(\citealt{Zhang2018NatCo...9..447Z}, their figure 3).
Hence there is no longer any statistical argument at hand
that points to a potentially much higher redshift of this burst. 

We conclude that the currently best strategy is to follow Occam's razor
and to adopt the hypothesis (\citealt{Rowlinson2010a})
that GRB 080905A was associated with the
large barred spiral at $z$=0.122. 

\section{Observations and Data Reduction}

\subsection{ATCA Radio-continuum Observations \label{ATCA}}

Radio-continuum observations of the host of GRB 080905A were performed
in two runs  several years after the burst on 2013 July 20  and on 2015
October 25 in the 5.5 and 9.0~GHz bands (corresponding to
wavelengths of 6 and 3~cm, respectively) with the Australia
  Telescope Compact Array (ATCA). The total combined observing time
was 10.5 hr (1st run: 1.51 hr, 2nd run: 9.04 hr). Both
observing runs were executed using the upgraded Compact Array
  Broadband Backend (CABB) detector (\citealt{Wilson2011}) and all
six 22 m antennae with the 6~km baseline (configuration 6A; program
ID: C2840, PI: A. Nicuesa Guelbenzu). In both runs, phase calibration
was performed by observing the quasar PKSB\footnote{Parkes Radio
Catalog in B1950 coordinates} 1908--201 (redshift $z=1.119$) for
3~minutes every hour followed by 57~minute integrations on target. This
quasar has  a radio flux of $F_\nu$(5.5~GHz) = 2.79~Jy and
$F_\nu$(9.0~GHz)=3.0~Jy.

Data reduction was performed using  the
Multichannel Image Reconstruction, Image Analysis and Display
(MIRIAD)\footnote{\url{http://www.atnf.csiro.au/computing/software/miriad/}}
software package for ATCA radio interferometry (for details, see
\citealt{Sault1995}). The Briggs ''robust'' parameter \citep{Briggs1995}
was varied between 0.0 and 2.0. The resulting 1~$\sigma_{\rm rms}$ of
the deconvolved images was between 5.4  and $5.7~\mu$Jy  beam$^{-1}$
at 5.5~GHz and between 6.5 and $7.4~\mu$Jy  beam$^{-1}$ at 9.0~GHz.
We  selected the results that gave the best compromise between
sensitivity and resolution (robust = 0.5). The width of the
synthesized beam (major axis, minor axis)
then was 7\farcs9$\,\times\,$1\farcs4 at 5.5~GHz and
4\farcs8$\,\times\,$0\farcs9 at 9.0~GHz.

\begin{figure}[t!]
\includegraphics[width=0.47\textwidth]{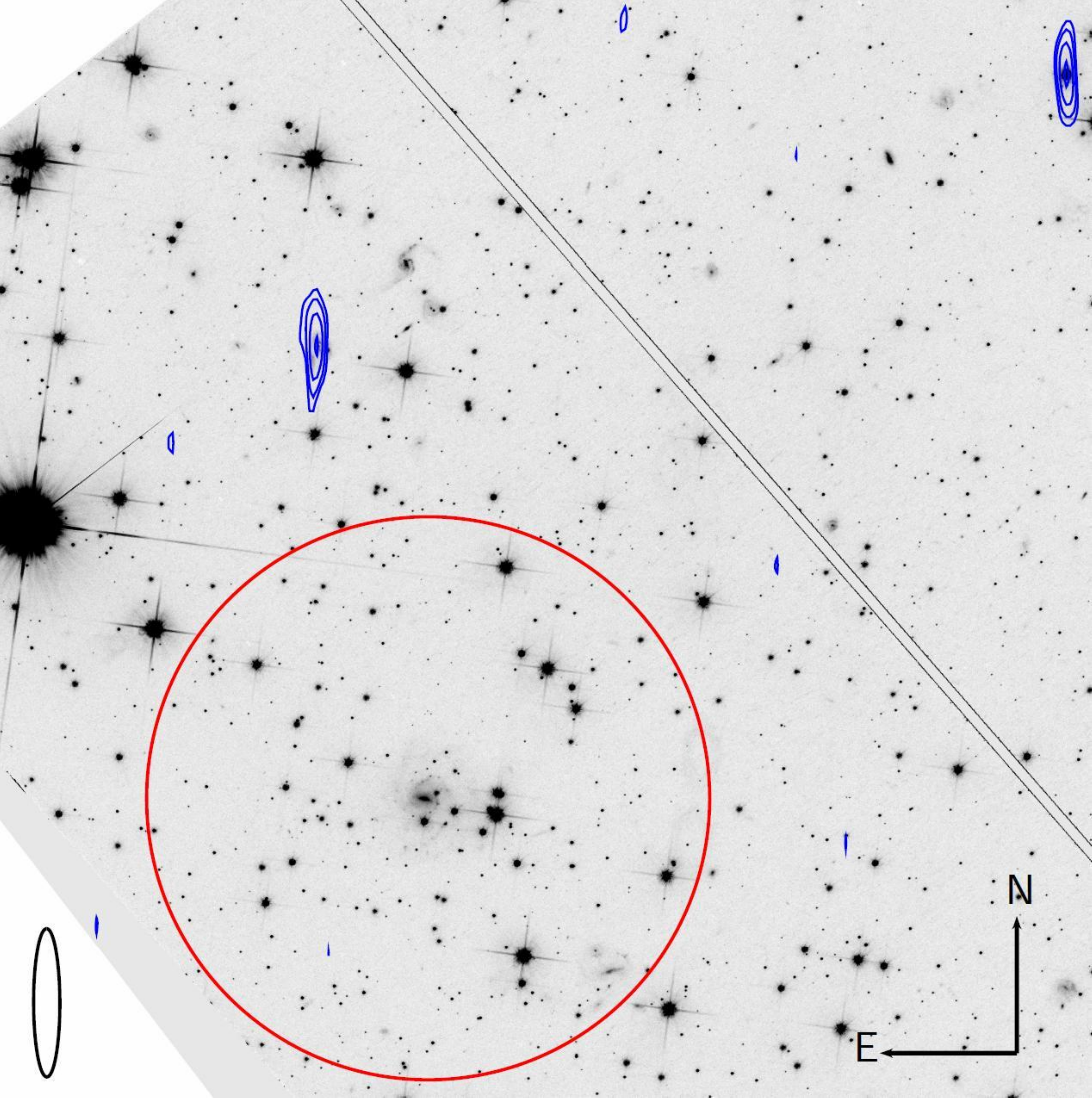}
\caption{
The field of GRB 080905A (field of view 2$'\times2'$) as seen by ATCA
in the 5.5~GHz radio-continuum band, overplotted the HST/WCS3 F814W
image taken on 2012 April 14 (program ID 12502; PI: A. Fruchter).
Radio contour levels (in blue) refer to $F_\nu$=20, 30, 50, and
80~$\mu$Jy.  To guide the eye, a circle with a radius of 30 arcsec
centered at the suspected GRB host galaxy is drawn (red color). The
radio source closest to the host ($F_\nu$(5.5~GHz) =
92$\pm$11~$\mu$Jy) is centered at the position of a tight pair of
(interacting?) galaxies at (radio-) coordinates R.A., decl. (J2000) =
19:10:42.854, $-$18:52:06.39 ($\pm$0\farcs10, $\pm$0\farcs65, resp.).
The radio source in the NW corner ($F_\nu$(5.5~GHz) =
91$\pm$8~$\mu$Jy) is an anonymous quasi-stellar object at (radio-)
coordinates R.A., decl. (J2000) = 19:10:37.217, $-$18:51:37.62
($\pm$0\farcs07, $\pm$0\farcs39, resp.).  The beam size and
orientation is shown in the lower left corner.}
\label{fig:ATCA}
\end{figure}

Despite the relatively small redshift, no radio source was detected superimposed on
the GRB host galaxy (Fig.~\ref{fig:ATCA}). The measured $3\sigma$ upper limits per beam are
$F_\nu($5.5~GHz$) < 17~\mu$Jy and $F_\nu($9.0~GHz$) < 20~\mu$Jy.

\subsection{VLT/MUSE Observations \label{MUSE}}

Observations with VLT/MUSE were performed on 30 May 2017 (program
ID: 099.D-0115A, PI: T. Kr\"uhler).  Eight dithered exposures of
$\sim$700~s each were obtained. Observations were executed using the
wide-field mode in which MUSE offers a field of view of $1'\,\times\,1'$.
In this mode, MUSE provides a pixel (spaxel) resolution of
0\farcs2. The MUSE data cover the wavelength range from 480 to
930~nm with a resolving  power of 1770 (480 nm)
- 3590 (930 nm).\footnote{\url{https://www.eso.org/sci/facilities/paranal/instruments/muse/overview.html}}
During the observations, the seeing 
was between 0\farcs9 (at 9000~\AA) and 1\farcs1 (at 5000~\AA).

The data were reduced following the methodology of
\cite{Kruehler2017}, using version 1.2.1 of the MUSE data reduction
pipeline provided by ESO
(\citealt{Weilbacher2012SPIE,Weilbacher2014ASPC}). The data were
corrected for Galactic foreground reddening ($E(B-V)$=0.12 mag;
\citealt{Schlafly2011}), assuming an average Milky Way extinction law
(\citealt{Pei1992}) and $R_V=3.08$. For the flux calibration, the
spectrophotometric standard star LTT3218 was observed at the beginning
of the night.

\begin{figure}[t!]
\includegraphics[width=0.48\textwidth]{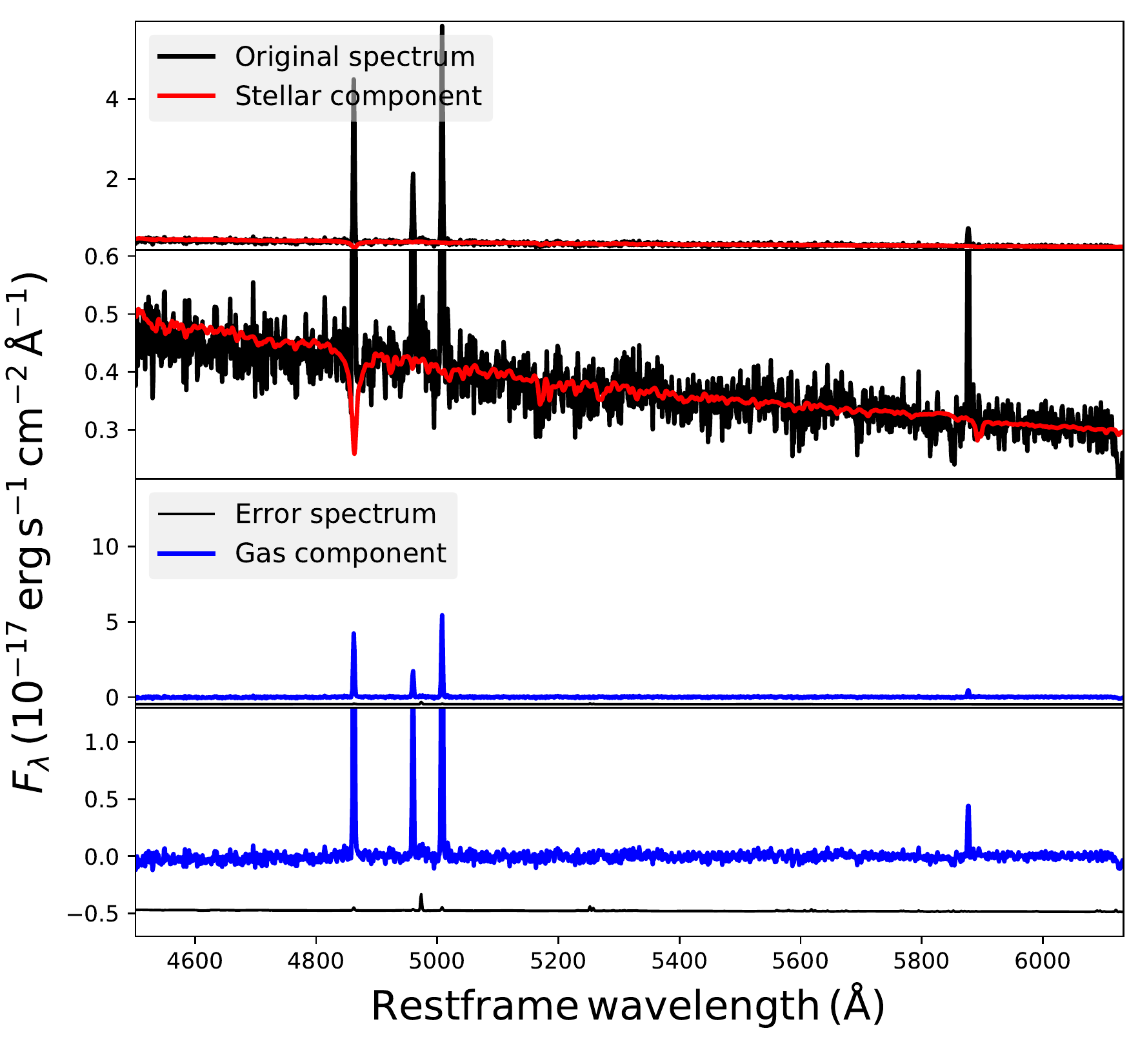}
\caption{ Spectrum and best-fit Starlight stellar continuum model
  corresponding to the 2x2 binned spectrum centered at the position of
  the HII region labeled A in Fig. 1, illustrating the separation of
  the stellar and gas-phase components. Top: original spectrum
  containing stellar and ionized gas emission (black) and our
  Starlight stellar continuum model fit (red). Top-middle:
  zoom-in of the continuum shown in the top panel.
  Bottom-middle: spectrum of the gas-phase-only component (blue)
  with the error spectrum indicated below in black. Bottom:
  zoom-in of the gas-phase-only component (blue) with the error
  spectrum now more visible (black).}
\label{fig:StarGas} 
\end{figure}

Analogous to \cite{Nicuesa2020} and following \cite{Kruehler2017}, we
separated the stellar and gas-phase components of the galaxy in order
to obtain accurate line flux measurements. We used the
Starlight software package (\citealt{Cid2005,Cid2009}) to model the
stellar continuum using a combination of single stellar population
models (\citealt{Bruzual2003}) and then subtracted the fitted stellar
continuum model to obtain the gas-phase-only data cube
(Fig.~\ref{fig:StarGas}).  In the
following, we used this gas-phase cube only, 
except when we calculated the equivalent widths. In the latter case
we used the combined cube (star + gas).

\subsection{Astrometry}

In a first step, the HST image was aligned using saoimage DS9
version 8.1 (\citealt{Joye2003ASPC..295..489J}) using The Fifth
  USNO CCD Astrograph Catalog, which provides positions with 10 to 70
mas precision (\citealt{Zacharias2017AJ....153..166Z}). We
double-checked the astrometric quality of the aligned image via a
bright, anonymous, radio point source in our ATCA 5.5~GHz images at
coordinates R.A., decl. (J2000) = 19:10:37.217, $-$18:51:37.615
($\pm$0\farcs05; $F_\nu$(5.5~GHz) = 91$\pm$8 $\mu$Jy).  For the
relatively bright optical counterpart of this  source (a spiral galaxy
seen face-on) we measured on the HST image central
coordinates R.A., decl. (J2000) = 19:10:37.221, $-$18:51:37.46
($\pm$0\farcs05). Based on this procedure, we estimate the accuracy of our
astrometry on the HST image to $\pm$0\farcs15. In a second
step, we aligned the MUSE data cube with the HST image via
stars visible in the  HST image close to the target position.
We estimate that our achieved relative astrometric accuracy
between HST and MUSE is $\pm0\farcs2$ ($\pm$1 spaxel) in each
direction.

\section{Results \label{Results} }

\subsection{Emission-line Fluxes and Their Uncertainties}

In order to measure the line fluxes, the numerical procedure
(\citealt{Kruehler2017,Tanga2018}) fits a Gaussian to an observed
emission line. Errors in the calculated line flux per spaxel are a
direct result of this fit. In addition, there are errors due to
uncertainties in the continuum level at the position of an emission
line (for a discussion of this issue, see also
\citealt{Erroz2019MNRAS.484.5009E}).
In our spectra, this mainly affects the \Hb \ line and becomes
evident by an apparently unphysical low flux ratio \Ha/\Hb\
(Sect.~\ref{EBV}).

In the following, when measuring a physical parameter for each
emission line we always adopted a threshold for the signal-to-noise ratio
(S/N). Consequently, if the flux in more than one emission line has to
be used in order to calculate a certain physical quantity, the line
with the smallest S/N determines if a certain spaxel enters the
statistics or not. The results obtained in this way are shown in the
following maps (see below).

In Appendix A we provide for all star-forming regions the
measured luminosity in the five lines \Ha, \Hb, \OIII 5007, \SII 6718, 
and \NII
6584.\footnote{According to the NIST Atomic Spectra Database, version 5.8,
at \url{https://www.nist.gov/pml/nist-atomic-spectra-bibliographic-databases},
in air, the line is centered at a laboratory wavelength of  6583.45~\AA\
(\citealt{NISTASD}). In the literature it is referred to as either \NII 6583 or \NII 6584.}
All spaxels with S/N$\geq$2 are used here. The number of spaxels
that fulfills this criterion differs from line to line.  Therefore,
in general, these emission-line luminosities should not be used to
calculate for a certain region an average $E(B-V)$, an average
extinction-corrected SFR, or an average metallicity index
(see also Sect.~\ref{meanvalues}).

\subsection{Radial Velocity Pattern and Dispersion Map}

\begin{figure}[t!]
\includegraphics[width=0.53\textwidth]{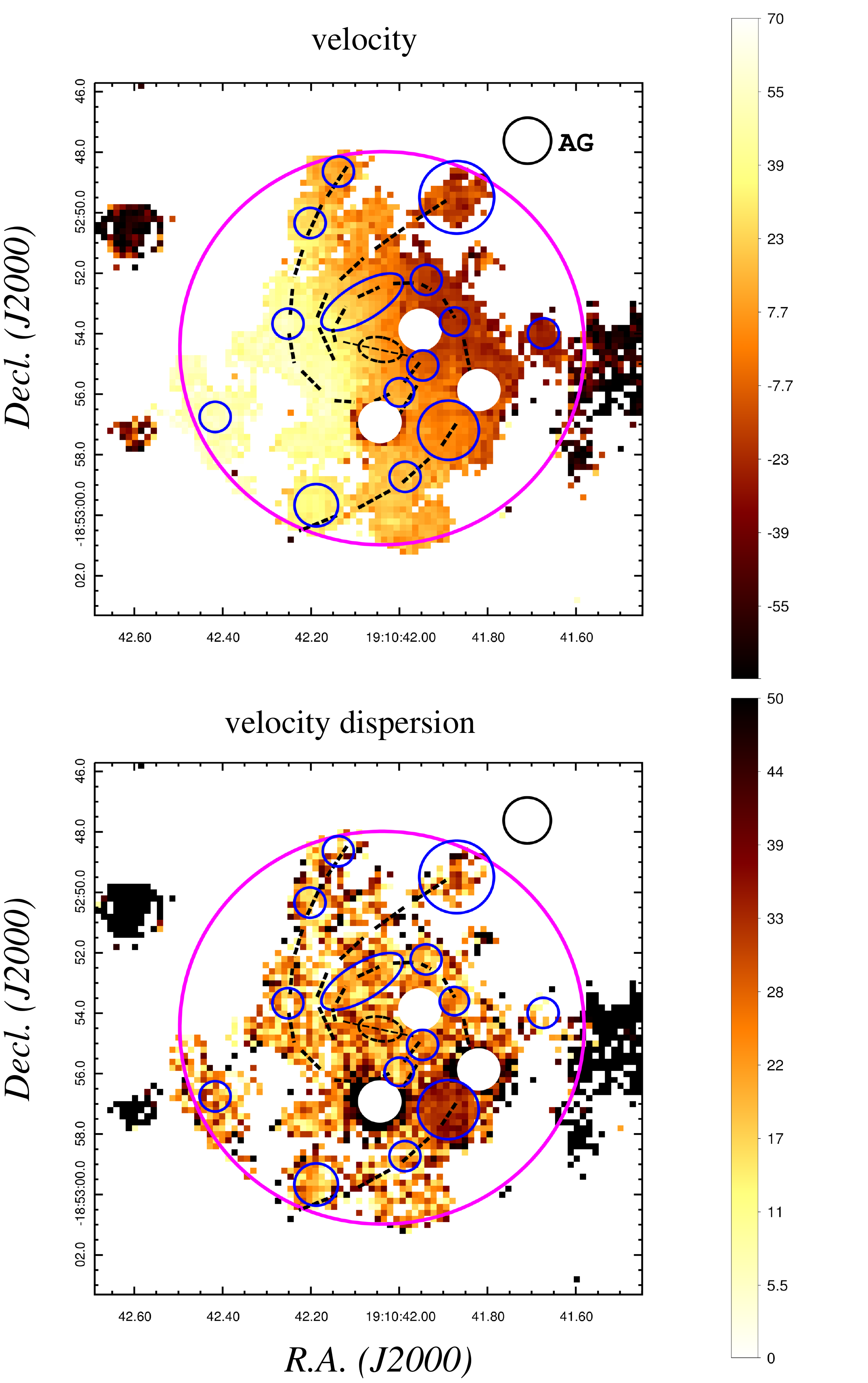}
\caption{Radial velocity (top) and velocity dispersion map (bottom) of
  the \Ha\ emission line of  the host of GRB 080905A.  One
  spaxel ($0\farcs2\,\times\,0\farcs2$) corresponds to
  $0.45\,\times\,0.45$~kpc$^2$. Velocities are given in units of km
  s$^{-1}$; velocities are observed, not corrected for inclination.
  The circles have the same meaning as in
  Fig.~\ref{fig:HST_080905}. The signals from the three bright
  Galactic foreground stars inside the big circle (magenta) have
  been masked.  Residuals from other bright Galactic foreground
  stars outside this circle have not been masked, however.}
\label{fig:vel}
\end{figure}

The radial velocity map shown in Fig.~\ref{fig:vel} is based on the
H$\alpha$ emission line, where  only spaxels with a signal-to-noise
ratio S/N$\geq$5 are plotted. H$\alpha$-emitting gas can be traced up to
about 7\farcs3 (16.5 kpc) away from the central bar of the galaxy.

The main feature in Fig.~\ref{fig:vel} is a velocity gradient from NW
to SE direction.  The velocity map confirms that the galaxy is not
seen exactly face-on.  According to \cite{Rowlinson2010a}, the
inclination angle is about 23 deg.

A comparison between the HST image (Fig.~\ref{fig:HST_080905}) and the
\Ha\ velocity map reveals a spatial asymmetry in the
\Ha\ emission component. MUSE detects a substantial amount of
\Ha-emitting gas in the eastern part of the galaxy around region \#10,
which has no counterpart on the western site of the galaxy. A possible
explanation could be that this is interstellar gas stripped from the
galaxy's disk due to an interaction with a nearby galaxy. However,
such a potential galaxy perturber cannot be identified in our data.

An additional clue about the \Ha-emitting gas comes from the velocity
dispersion map (Fig.~\ref{fig:vel}, bottom). When calculating
this map we adopted $10^4$~K warm gas ($\sigma_{\rm gas}$ =
9.1~km s$^{-1}$) and a MUSE line-spread function (LSF) of 2.5~\AA
\ at  the wavelength of the redshifted \Ha\ line (see Fig. 15 in
\citealt{Bacon2017A&A...608A...1B}), which  corresponds to an FWHM of
101~km s$^{-1}$ and a $\sigma_{\rm instr}$ of 43~km s$^{-1}$ (for
more details on our procedure, see \citealt{Nicuesa2020}).
Neglecting the artificial features that result from the removal of the
Galactic foreground stars, we find the following: the velocity
dispersion is highest in the bright star-forming region A, where in a
circular area with a radius of 1\farcs2 (2.7 kpc) the velocity
dispersion has a mean of 34 km s$^{-1}$.  Outside region A and
across the entire galaxy, the dispersion lies between about 10 and 30
km s$^{-1}$, with no additional peak analogous to region A. In
particular, region \#10 does not stand apart in any way. Region X (in
projection closest to the GRB explosion site) is clearly
detected with a median dispersion of 26 km s$^{-1}$.\footnote{In
Fig.~\ref{fig:vel} regions with apparently substantially higher
dispersion velocities are due to imperfectly removed Galactic
foreground stars.}

In any case, measuring the galaxy's kinematics by using only
the \Ha\ emission line is not the most accurate way.  MUSE is an
electronic detector using pixels and as a consequence of that, the MUSE
line-spread function is undersampled. Pixellation affects the random
noise errors in wavelength (\citealt{Robertson2017PASA...34...35R})
and adds an additional measurement error to the deduced velocity
field. Therefore, a more accurate determination of the kinematics of
the gas should make use of several emission lines (e.g.,
\citealt{McLeod2015MNRAS.450.1057M}). However, given that our
scientific focus is not mainly kinematics, we do not
investigate this further.


\begin{deluxetable*}{rcccrr }
\tablewidth{0pt} 
\tablecaption{Star-forming Regions in the Host of GRB 080905A.}
\tablehead{
  \colhead{Region    }   &  
  \colhead{R.A., Decl.}  &  
  \colhead{Radius  }     &  
  \colhead{Vel. disp.}   &  
  \colhead{EW(\Ha)}      &      
  \colhead{SFR(\Ha)}  \\      
  \colhead{        }     &  
  \colhead{19:10:, $-$18:52:}  &  
  \colhead{arscec  }     & 
  \colhead{km s$^{-1}$}  & 
  \colhead{\AA}          & 
  \colhead{0.01~\msunyr}      } 
\startdata
    A  & 41.89, 57.2 &  1.0       & 33.7 $\pm$ 0.4 &   80.6 $\pm$ 0.2  & 24.27 $\pm$ 0.05\\
    B  & 42.19, 59.7 &  0.7       & 20.5 $\pm$ 1.3 &  109.0 $\pm$ 0.8  &  4.22 $\pm$ 0.03\\
    D  & 42.08, 52.9 &  1.5 x 0.6 & 20.6 $\pm$ 0.8 &   31.2 $\pm$ 0.2  &  7.73 $\pm$ 0.04\\[1mm]
    1  & 41.67, 54.0 &  0.5       & 15.2 $\pm$ 2.5 &   27.8 $\pm$ 0.6  &  0.88 $\pm$ 0.02\\
    2  & 41.87, 53.6 &  0.5       & 22.3 $\pm$ 1.1 &   27.0 $\pm$ 0.3  &  2.29 $\pm$ 0.02\\
    3  & 41.94, 52.2 &  0.5       & 18.9 $\pm$ 1.1 &   40.5 $\pm$ 0.5  &  1.66 $\pm$ 0.02\\
    4  & 41.95, 55.0 &  0.5       & 22.8 $\pm$ 1.2 &   26.3 $\pm$ 0.2  &  3.72 $\pm$ 0.03\\
    5  & 41.99, 58.7 &  0.5       & 21.6 $\pm$ 1.7 &   45.1 $\pm$ 0.7  &  1.48 $\pm$ 0.02\\
    6  & 42.00, 55.9 &  0.5       & 20.6 $\pm$ 1.4 &   18.0 $\pm$ 0.2  &  3.30 $\pm$ 0.03\\
    7  & 42.14, 48.6 &  0.5       & 22.2 $\pm$ 3.3 &   58.6 $\pm$ 1.8  &  0.95 $\pm$ 0.02\\
    8  & 42.20, 50.3 &  0.5       & 19.3 $\pm$ 1.8 &   67.6 $\pm$ 1.1  &  1.36 $\pm$ 0.02\\
    9  & 42.25, 53.7 &  0.5       & 21.0 $\pm$ 1.6 &   41.1 $\pm$ 0.7  &  1.15 $\pm$ 0.02\\
   10  & 42.42, 56.7 &  0.5       & 23.8 $\pm$ 2.0 &   56.9 $\pm$ 1.0  &  1.30 $\pm$ 0.02\\
\hline                                                                                    
bar; C & 42.04, 54.5 &  0.7 x 0.4 & 24.5 $\pm$ 1.1 &    8.5 $\pm$ 0.1  &  2.58 $\pm$ 0.03\\
    X  & 41.87, 49.5 &  1.2       & 26.3 $\pm$ 1.6 &   19.4 $\pm$ 0.4  &  2.82 $\pm$ 0.04\\
galaxy & 42.04, 54.5 &  6.5       & 27.3 $\pm$ 0.7 &   29.9 $\pm$ 0.1  &141.77 $\pm$ 0.25\\ 
\enddata
\tablecomments{
Coordinates refer to the centers of the circles drawn in Fig.~\ref{fig:HST_080905}.
In the case of C and D the minor and major axes of the drawn ellipse are given.
A, B, and D are the most luminous star-forming regions in the galaxy. The list
of the less luminous regions is not complete. The last column gives SFR(\Ha)
based on the measured \Ha \ luminosity (Appendix A,
Eq.~\ref{SFRHa}), not corrected for internal host-galaxy extinction.
The results obtained for regions \#2, 4, and 6 have to be taken with care since
these regions lie close to two bright foreground stars (Fig.~\ref{fig:HST_080905}).
After the gas-star separation in the MUSE data cube (Sect.~\ref{MUSE})
these stars show detectable residual emission extending into these regions.
}
\label{Tab:A1}
\end{deluxetable*}

\subsection{Host-galaxy Reddening $E(B-V)$ \label{EBV}}

\subsubsection{Procedure}

For each spaxel,  we calculated the internal host-galaxy reddening via
the Balmer decrement (e.g., \citealt{Dominguez2013ApJ...763..145D}),
adopting the standard approach (case B
recombination at $T=10^4$~K, electron density $10^2-10^4$~cm$^{-3}$;
\citealt{Osterbrock1989}). In doing so, we assumed a Milky Way
extinction law (\citealt{Pei1992})
with a ratio of total-to-selective extinction $R_{\rm
V}$ = 3.08. Then
\begin{equation}
E(B-V)_{\rm host} = 1.98 \, \log_{10}((\mbox{\Ha/\Hb})/2.86)\,, 
\label{Balmer}
\end{equation}
where H$\alpha$/H$\beta$ is the observed flux ratio in the lines.  In
our calculations we required that S/N$\geq$2 for both lines
simultaneously. The results obtained in this way are shown in
Fig.~\ref{fig:ebv}. In this plot, spaxels for which we found an
unphysical $E(B-V)<0$ were set to $E(B-V)=0$ (e.g.,
\citealt{Reddy2015ApJ...806..259R,Erroz2019MNRAS.484.5009E}). The reddening map
nicely tracks the spiral pattern of the galaxy. The reddening is highest
in the galaxy's bar and lowest in the star-forming region A.

\subsubsection{The Issue of Apparently Negative Reddening Values}

Calculated negative reddening values are unphysical, which raises the
question of what 
their origin could be. \cite{Jimmy2016ApJ...825...34J} have addressed
this issue already in greater detail, so that we just briefly focus on
the case discussed here.

Equation~\ref{Balmer} relies on the assumption of a Milky Way
extinction law (e.g., \citealt{Dominguez2013ApJ...763..145D}).
Other extinction laws will lead to a factor different from 1.98 (e.g.,
\citealt{Reddy2015ApJ...806..259R, Baron2018MNRAS.480.3993B}), but will
not decrease the expected \Ha\ to \Hb\ flux ratio.

In principle, the expected \Ha/\Hb\ flux ratio can be reduced
by changing the temperature and spatial density of the gas away from
the standard assumptions (\citealt{Osterbrock1989}). However,
the \Ha/\Hb\ flux ratio is not very sensitive to such changes;
for a wide parameter range, the \Ha/\Hb\ flux ratio 
does not drop below about 2.7 (e.g., \citealt{Caplan1986A&A...155..297C,
Groves2012MNRAS.419.1402G, Li2019ApJ...872...63L}).

We found two possible explanations for calculated
negative reddening values. (i) Spaxels close to the position of the
bright Galactic foreground stars are affected by a poor
continuum subtraction (for an analogous discussion, see
\citealt{Erroz2019MNRAS.484.5009E}). The \Ha\ and \Hb\
lines at these positions are clearly not Gaussians, which affects the
numerical determination of the flux in the lines. (ii) Several other spaxels
have a low S/N in the lines; the 
line fluxes are not measured at high significance, finally leading to
an apparent \Ha/\Hb \ flux ratio of less than 2.86.

\subsubsection{Calculating Mean Values \label{meanvalues} }

\begin{figure}[t!]
\includegraphics[width=0.53\textwidth]{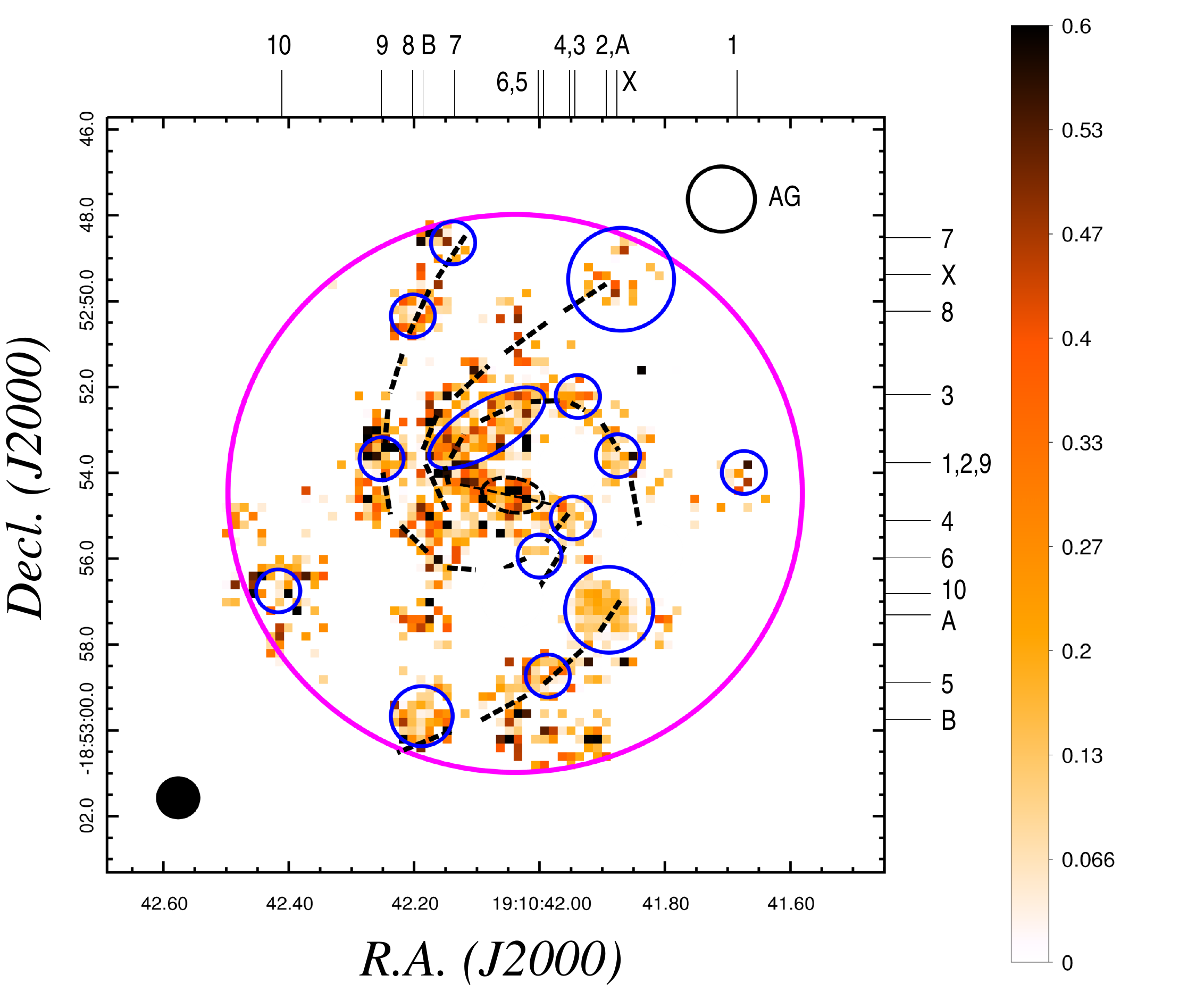}
\caption{Map of the internal host-galaxy reddening $E(B-V)$ (in
  magnitudes) calculated via the measured Balmer decrement. The
  circles have the same meaning as in Figs.~\ref{fig:HST_080905} and
  \ref{fig:vel}. Also shown is the seeing disk (diameter 1
  arcsec).}
\label{fig:ebv}
\end{figure}

In order to calculate the mean reddening in an individual star-forming
region, we considered two approaches. (i) In a first approach, we started
with the total flux in the \Ha\ and \Hb\ lines
(Appendix A) and then calculated the corresponding mean
$E(B-V)$ via Eq.~\ref{Balmer}. A shortcoming of this method is that
for a required S/N  the number of spaxels that can be used to measure
the total flux can be different from line to line. Moreover, for the
case discussed here, due to residual light from the Galactic foreground
stars, in some regions more spaxels had to be masked out in the \Hb \
line than in the \Ha \ line. (ii) In a second approach, we performed 
statistics based on the corresponding reddening values for all
individual spaxels (Fig.~\ref{fig:ebv}). In doing so, spaxels with
$E(B-V)<0$ were not included in the calculation of the mean.  A
shortcoming of this method is that it introduces a bias in the
statistics if spaxels with $E(B-V)<0$ are ignored.

A comparison of the results obtained
via both methods showed a good agreement in the case of the \Ha-brightest
star-forming regions, i.e., those regions where the spaxels have
the highest S/N in the lines (A, B, and D). For the less luminous
regions, however, this agreement was not as good. Not surprisingly,
these regions are characterized by a low S/N in at least one line.

Since a detailed study of all star-forming
regions in the host of GRB 080905A is beyond the scope of this paper,
we finally did not consider this issue further and focused only
on the \Ha-brightest star-forming regions (Table~\ref{Tab:A2}).

\subsection{Star formation rate and star-forming regions \label{SFR} }

Star formation rates were calculated via the extinction-corrected
H$\alpha$ emission-line flux following \cite{Murphy2011ApJ...737...67M}
who used {\tt Starburst99} (\citealt{Leitherer1999ApJS..123....3L}) and a
Kroupa IMF (\citealt{Kroupa2001MNRAS.322..231K}).
According to these authors,
\begin{eqnarray}
\mbox{SFR(H}\alpha) = 5.37\times10^{-42}\
L(\rm H\alpha) \ M_\odot\ \mbox{yr}^{-1}\,,
\label{SFRHa}
\end{eqnarray}
where the \Ha\ luminosity $L$ is measured in units of erg\, s$^{-1}$.

\begin{figure}[t!]
\includegraphics[width=0.53\textwidth]{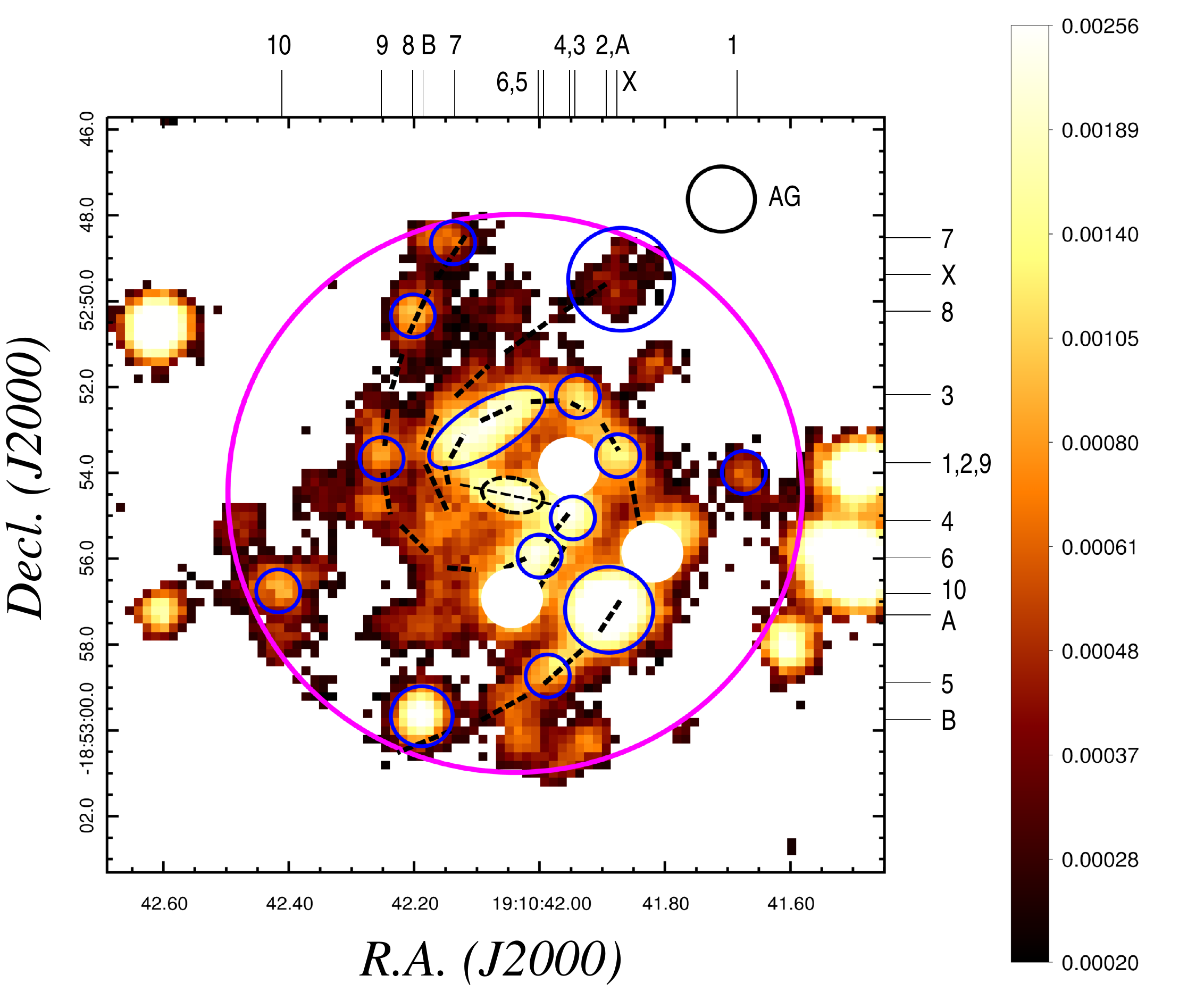}
\caption{Map of the extinction-corrected star formation rate
  (for each spaxel in units of $M_\odot$\,yr$^{-1}$). The
  circles have the same meaning as in Figs.~\ref{fig:HST_080905} and
  \ref{fig:vel}. Compared to Fig.~\ref{fig:ebv}, here also
  regions with an adopted $E(B-V)=0$ are plotted.  Similar to
  Fig.~\ref{fig:vel}, outside the larger circle (in magenta),
  residual flux from bright Galactic foreground stars is seen.
  Only spaxels with an S/N$\geq5$ in the measured SFR are plotted here.
}  
\label{fig:sfr}
\end{figure}

\begin{figure}[t!]
\includegraphics[width=0.48\textwidth]{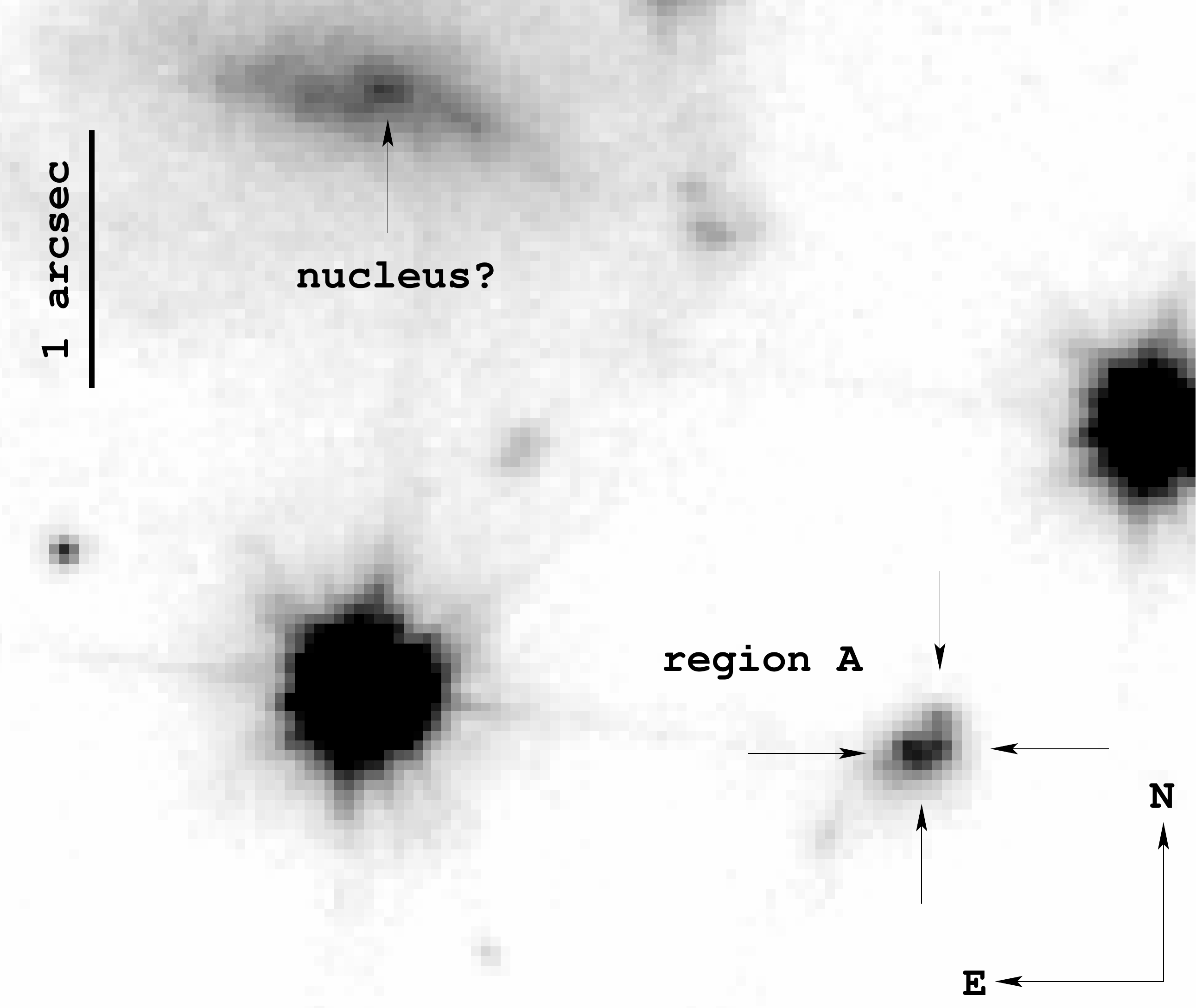}
\caption {Zoom-in into the HST/F606W image (Fig.~\ref{fig:HST_080905})
  at the position of the H$\alpha$-bright star-forming region A.  It
  consists of at least four individual bright knots, tightly packed
  together within an area with a diameter of  0\farcs3-0\farcs4
  ($\sim$0.7-0.9 kpc). Also indicated is a bright spot in the central
  bar of the galaxy, probably the luminous galactic nucleus.}
\label{fig:HST_zoom}
\end{figure}

The SFR map reveals several star-forming regions close to the central
bar (see also Fig.~\ref{fig:HST_zoom}) and scattered along the spiral
arms up to about 15 kpc distance from the center of the galaxy
(Fig.~\ref{fig:sfr}). All these regions can also be identified on the
\HST\ image (Fig.~\ref{fig:HST_080905}).  In particular, the outer
south-western arm pops up as a place with a high star formation
activity (region A), in contrast to the outer northeastern arm, which
shows comparably weak star formation activity (regions \#7-9).

The \Ha-brightest star-forming region (designated A) 
lies about 10 arcsec (22 kpc) away from the afterglow position and
about 3\farcs4 (7.6 kpc) away from the galaxy's central
bar. It has a bright, morphologically resolved counterpart on the
HST/F606W image (Fig.~\ref{fig:HST_zoom}). For this region, we measure a
H$\alpha$ luminosity of about $5\,\times\,10^{40}$ erg s$^{-1}$
within a circle with a radius of 1\farcs0 (2.25 kpc).
This corresponds to an extinction-corrected 
SFR of about 0.30~$M_\odot$\,yr$^{-1}$ (Table~\ref{Tab:A2}).

When compared to the other star forming regions (Fig.~\ref{fig:ebv}),
region A shows the lowest amount of reddening by dust.
A trend of a decreasing reddening with increasing \Ha \ luminosity
has been found in star-forming regions in
other galaxies and supports a scenario in which the dust is
destroyed or swept up by the intense radiation field of the most massive stars
(e.g., \citealt{Cairos2017A&A...600A.125C}).

\begin{figure*}[ht!]
\includegraphics[width=1.0\textwidth]{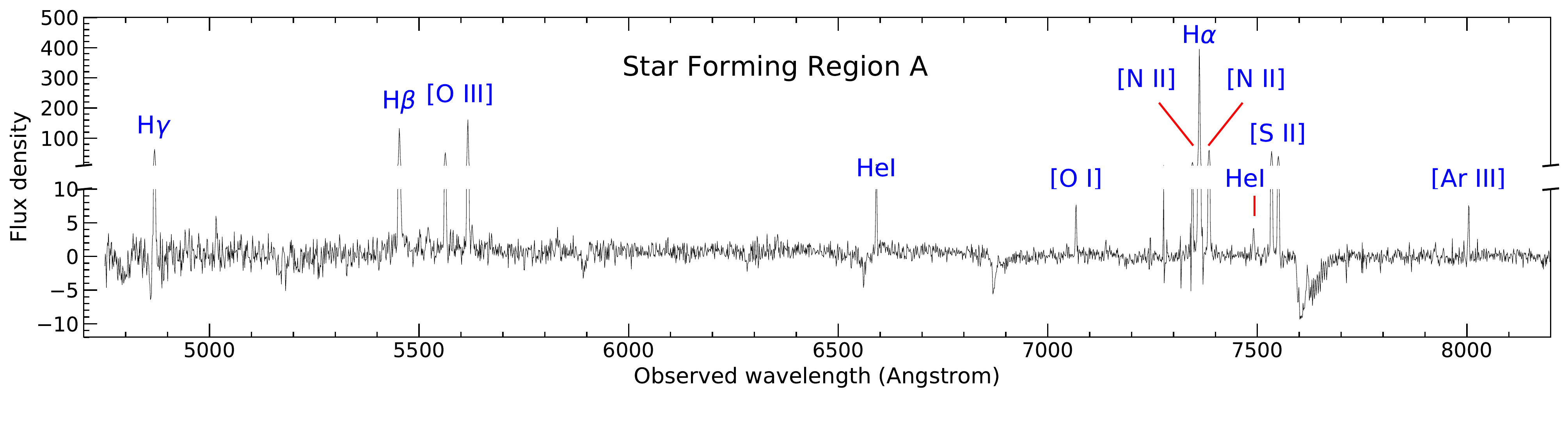}
\caption{MUSE spectrum (gas cube; Sect.~\ref{MUSE})  of the most
  luminous star forming region (A) in the GRB 080905A host galaxy.
  The flux density is given in units of $10^{-20}$ erg s$^{-1}$
  cm$^{-2}$ \AA $^{-1}$  The most prominent emission lines are
  indicated: [O\,{\sc iii}] stands for the lines at 4958.9 and
  5006.8~\AA, [N\,{\sc ii}] for the lines at 6548.1 and 6583.4~\AA,  
  and [S\,{\sc ii}] for the lines at 6716.4 and 6730.8~\AA.
  The absorption features near 6550, 6850, and
  7600~\AA\ are of telluric origin. Note that we have split the
  spectrum into two parts in order to show the peaks of the bright and
  the faint emission lines.}
\label{fig:Spectra}
\end{figure*}

Figure~\ref{fig:Spectra} shows the MUSE spectrum of region A,
averaged over a region with a radius of 1$''$ (79 spaxels).
Identified emission lines are labeled (for the laboratory wavelengths in air
see \citealt{NISTASD}).

No obvious \Ha\ line emission is detected from the area that contains
(in projection) the GRB explosion site (black circle in
Fig.~\ref{fig:sfr}). The GRB progenitor exploded in an environment
that is apparently less filled with line-emitting gas.
In this area, we find SFR(\Ha)$<$0.003 \msunyr (3$\sigma$), measured
in a region with the diameter of the seeing disk (1 arcsec).

The \Ha \ emission region that is closest to the position of the optical
afterglow is region X, an isolated island (black circle in
Fig.~\ref{fig:HST_080905}) at central coordinates R.A., decl. (J2000)
= 19:10:41.87, $-$18:52:49.5. It lies about 2\farcs9 (6.5 kpc) away
from the center of the afterglow error circle.  Line emission from its
outskirts can be followed up to a distance as close as 1\farcs7 (3.8
kpc) to the central afterglow position (\citealt{Rowlinson2010a}).
Region X shows a H$\alpha$ luminosity that corresponds to an SFR of
$\sim$0.03~$M_\odot$\,yr$^{-1}$ (Table~\ref{Tab:A1})
within a circle with a radius of 1\farcs2
(2.7 kpc). Contrary to the star-forming regions listed in
Table~\ref{Tab:A1}, this is an area of diffuse H$\alpha$ emission
with no centrally dominant peak.

For the entire galaxy we find a H$\alpha$ luminosity of about
$2.6\,\times\,10^{41}$ erg s$^{-1}$ (Appendix A),
corresponding to an extinction-corrected SFR of
about 1.6~$M_\odot$\,yr$^{-1}$ inside a circle with a radius of 6\farcs5
(15 kpc, Table~\ref{Tab:A2}) centered at the central
bar of the host. Given this SFR and the mass estimate from
\cite{Rowlinson2010a}, the specific SFR is about
1$\,\times\,10^{-10}$ yr$^{-1}$.  For short-GRB hosts, this is a
typical value (\citealt{Berger2014ARAA}).

An SFR(\Ha) of $\sim$1.6~$M_\odot$\,yr$^{-1}$
has to be compared with the upper
limit obtained from the ATCA observations (Sect.~\ref{ATCA}).
Following \cite{Greiner2016A&A593A}, the constraint on the SFR is
strongest for the 5.5~GHz data. For a Briggs robust parameter of 0.5,
this leads to an SFR(radio) $<$1.1~$M_\odot$\,yr$^{-1}$ per beam.  Given that
several beam sizes are required to cover the entire GRB 080905A host,
the radio upper limit is consistent with the \Ha-derived SFR.

\subsection{Metallicity \label{metallicty}}

\begin{figure}[t!]
\includegraphics[width=0.53\textwidth]{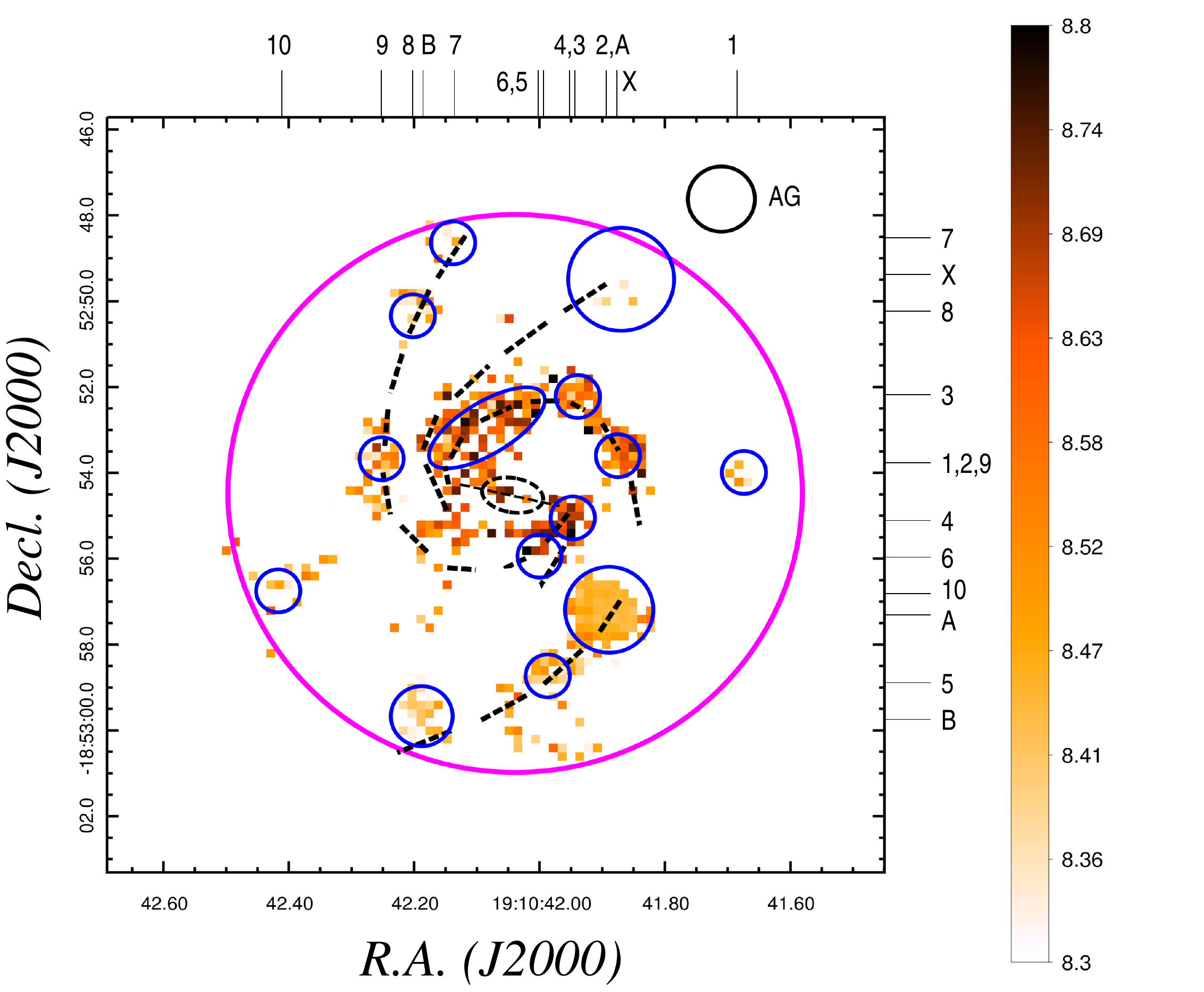}
\caption{Map of the O3N2 metallicity indicator, given as 12 +
  log(O/H). The circles have the same meaning as in
  Figs.~\ref{fig:HST_080905} and \ref{fig:vel}.
}
\label{fig:metal}
\end{figure}

In order to calculate the nebular oxygen abundance, we followed
\citet[PP04]{Pettini2004}, according to whom the nebular oxygen abundance
can be calculated as
\begin{eqnarray}
12+\log\mbox{(O/H)} &=& 8.73 - 0.32 \times \mbox{O3N2}\,, \,\,\, \mbox{with} \label{eq:O3N2}\\
\mbox{O3N2} &\equiv& \log \frac{ \mbox{\OIII} 5007 / \mbox{H}\beta}
{\mbox{\NII} 6584 / \mbox{H}\alpha}\,\nonumber\,.
\end{eqnarray}
In Fig.~\ref{fig:metal} we plot only spaxels for which S/N$\geq$2
simultaneously in \Ha, \OIII 5007, H$\beta$, and \NII 6584.
Spaxels for which the flux ratio \Ha/\Hb \ was smaller
than 2.86 are not plotted here.

Similarly to the reddening map, the metallicity map based on the PP04
formulation traces the spiral structure of the galaxy. What is
immediately apparent in Fig.~\ref{fig:metal} is the metallicity
gradient from the center of the galaxy to its outer regions.
In addition, the star-formation sites A and B reveal themselves as
large metal-poor regions (Table~\ref{Tab:A2}).
In region A the mean of 12+log(O/H) is
about 8.46, corresponding to $Z/Z_\odot$ = 0.59.\footnote{For a solar
value of 12+log(O/H) = 8.69 (\citealt{Asplund2009ARA&A..47..481A};
but see also \citealt{Kewley2019ARA&A..57..511K} and
\citealt{Vagnozzi2019Atoms...7...41V}).}

The metallicity is highest close
to the galaxy's bar, where 12+log(O/H) reaches the solar
value.  For the entire galaxy we measure a mean
of $8.52\pm0.01 \ (Z/Z_\odot = 0.68\pm0.01)$. These results
are in good agreement with VLT/FORS1 long-slit spectroscopy
(\citealt{Rowlinson2010a}). In particular, the apparent
north-south asymmetry in the metallicity of the galaxy found by
these authors turns out to be a consequence of the dominating nature
of the giant, metal-poor, star-forming region A in the southern spiral
arm.

\begin{deluxetable}{r rrr }
\tablewidth{0pt} 
\tablecaption{Calculated Mean Values for the Brightest Star-forming Regions.}
\tablehead{
  \colhead{region    }   &  
  \colhead{$E(B-V)$}     &  
  \colhead{SFR }         &  
  \colhead{Metallicity}  \\ 
  \colhead{        }     &  
  \colhead{mag     }     & 
  \colhead{0.01~\msunyr} & 
  \colhead{12+log(O/H)} } 
\startdata
    A  &  0.08$\pm$  0.01        & 31.0$\pm$1.4      &  8.46$\pm$ 0.01\\
    B  &  0.14$\pm$  0.06        &  6.5$\pm$1.2      &  8.36$\pm$ 0.03\\
    D  &  0.19$\pm$  0.05        & 13.8$\pm$2.0      &  8.67$\pm$ 0.02\\[1mm]
bar; C &  0.45$\pm$  0.11        & 10.2$\pm$3.3      &  8.62$\pm$ 0.05\\
    X  &  0.21$\pm$  0.15        &  5.4$\pm$2.6      &  8.39$\pm$ 0.07\\
galaxy &  0.04$\pm$  0.01        &160.8$\pm$7.0      &  8.52$\pm$ 0.01\\ 
\enddata
\tablecomments{ 
All but one value have been calculated using the measured total
emission-line luminosities listed in Appendix A.
The exceptional case is the mean metallicity in region C. Here
most spaxels have a very low S/N in the [O\,{\sc iii}] line.
Therefore, we calculated the mean using
those 6 spaxels for which S/N$>$2 is fulfilled in all four
lines (Fig.~\ref{fig:metal}).
For comparison, we also provide the corresponding numbers for the central bar (C)
and for the entire galaxy, after masking out residual flux from the
three bright Galactic foreground stars.
In addition, we give the corresponding numbers for the
region which in projection lies closest to the GRB explosion site (X).
We caution, however, that here all numbers have a relatively large error.
}
\label{Tab:A2}
\end{deluxetable}

\subsection{Equivalent Width of \Ha \label{EW} }

\begin{figure}[t!]
\includegraphics[width=0.53\textwidth]{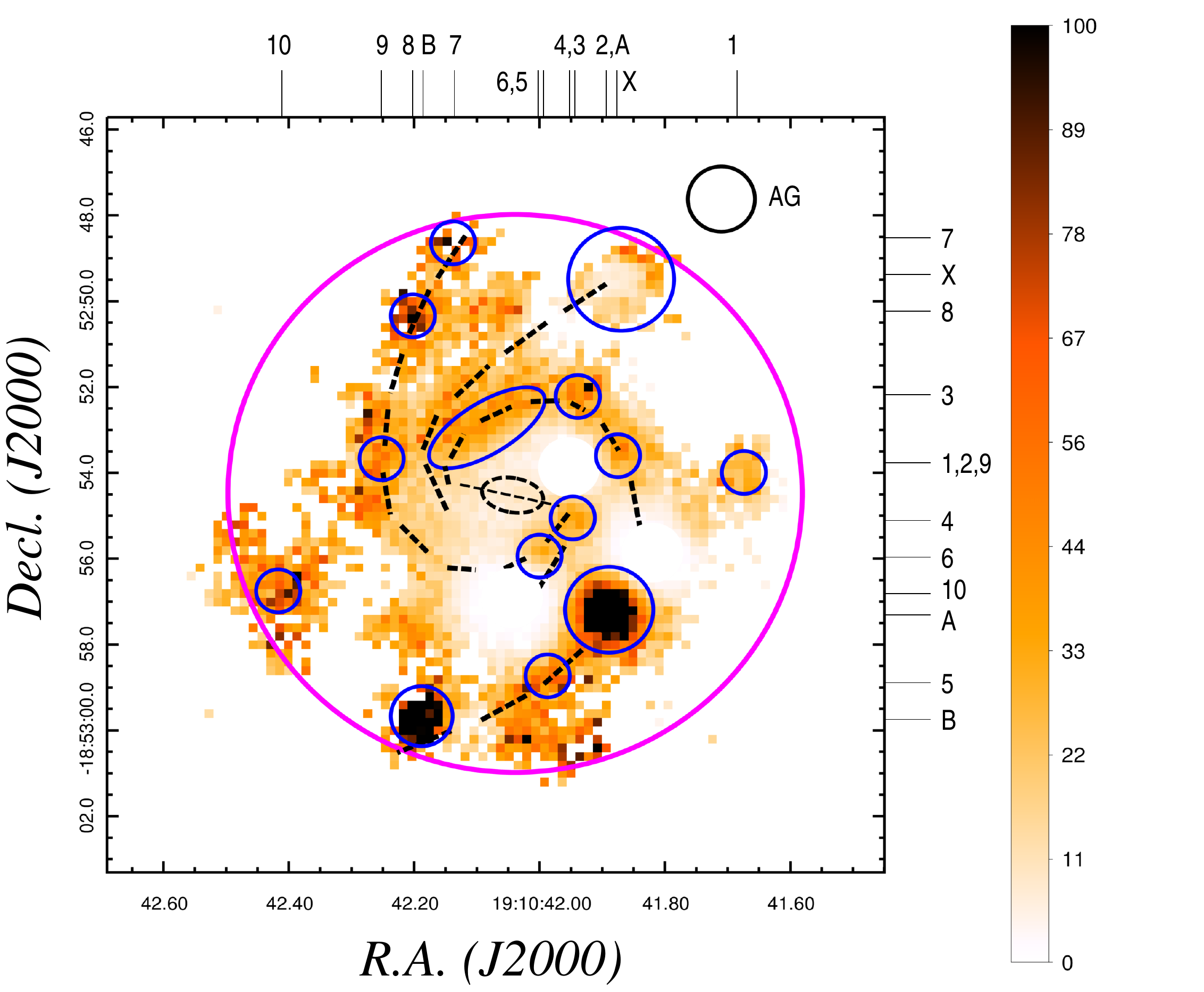}
\caption{Map of the equivalent width EW(\Ha) in units of \AA. The
  star-forming regions A and B stand out here because of their very
  high EW values.  The circles have the same meaning as in
  Figs.~\ref{fig:HST_080905} and \ref{fig:vel}.}
\label{fig:EW}
\end{figure}

The EW(\Ha) map was constructed using spaxels for which
S/N(\Ha)$\geq$4 (Fig.~\ref{fig:EW}). The map reveals a number of
interesting features. The equivalent width is lowest in the central
bar (C) of the galaxy (between 7 and 10~\AA). It is much higher along
the spiral arms (e.g., around 20--50~\AA \ in region D), and highest
in region A, where it peaks at about 200~\AA. In region X it shows a
larger valley (EW(\Ha) between 6 and 9~\AA) but increases to 30 --
50~\AA \ at its northern border. In regions \#8 and \#10 the scatter
in EW(\Ha) is relatively high, the median is 68 and 56~\AA,
respectively.  For the entire galaxy (within the bigger circle), we
measure a mean of 29 and a  median of 23~\AA\ (Table~\ref{Tab:A1}).

In order to interpret these results, we made use of the {\tt Starburst99}
synthesis model (\citealt{Leitherer1999ApJS..123....3L}).
Its publicly available database\footnote{\url{
https://www.stsci.edu/science/starburst99/docs/popmenu.html}}
provides a relationship between EW and the age of a star-forming region
for three different initial mass functions (IMFs) and five
metallicities between $Z$=0.040 and 0.001.
Using this database, for the metallicities of interest here
($Z$= 0.009 -- 0.020; Table~\ref{Tab:A2}), all star-forming
  regions have an age between 6 and 8 Myr; the results for
  different IMFs differ by at most 0.2 Myr.
Nevertheless, other stellar population models like BPASS
(\citealt{Eldridge2017PASA...34...58E}) can lead to older
age estimates (e.g., \citealt{Kuncarayakti2013AJ....146...30K,
Kuncarayakti2016A&A...593A..78K}). For example, 
for single stellar-evolution models, the age range for $Z$=0.009-0.020 for
EW(\Ha)=30-60~\AA\ is 8-12~Myr. In particular,
binary stellar-evolution models, which within the same metallicity and
EW range, give an age range of 20-40~Myr. We do not
investigate this further.  

The trigger of the recent star-forming activity in the suspected host
of GRB 080905A is still unclear.  As noted above, in the data
presented here we do not see evidence for any galaxy-galaxy
interaction. Future HI 21~cm or other atomic and molecular line
observations could be a promising tool to address this question (e.g.,
\citealt{Arabsalmani2015MNRAS.454L..51A,
  Michalowski2015A&A...582A..78M,
  Michalowski2016A&A...595A..72M,
  Michalowski2018A&A...617A.143M,  
  Michalowski2020A&A...642A..84M,
  Hatsukade2020ApJ...892...42H}).

\subsection{Emission-line Diagnostic Diagrams \label{BPT}}

We follow previous work (\citealt{Nicuesa2020}) and use diagnostic
Baldwin-Philips-Terlevich (BPT) emission-line diagrams (\citealt{BPT1981})
based on the line ratios  \OIII 5007/H$\beta$ vs. \NII 6584/H$\alpha$
and \OIII 5007/H$\beta$ vs. \SII 6718/H$\alpha$ to distinguish between
stellar ionization of the gas (\HII; star forming)  and
other ionization processes (stellar winds, AGN activity, shocks).
In Fig.~\ref{fig:metal} we plot only spaxels for which S/N$\geq$2
simultaneously in all four lines. Spaxels that include residual light from the three
Galactic foreground stars have been masked out here.

\begin{figure*}
\includegraphics[width=0.50\textwidth]{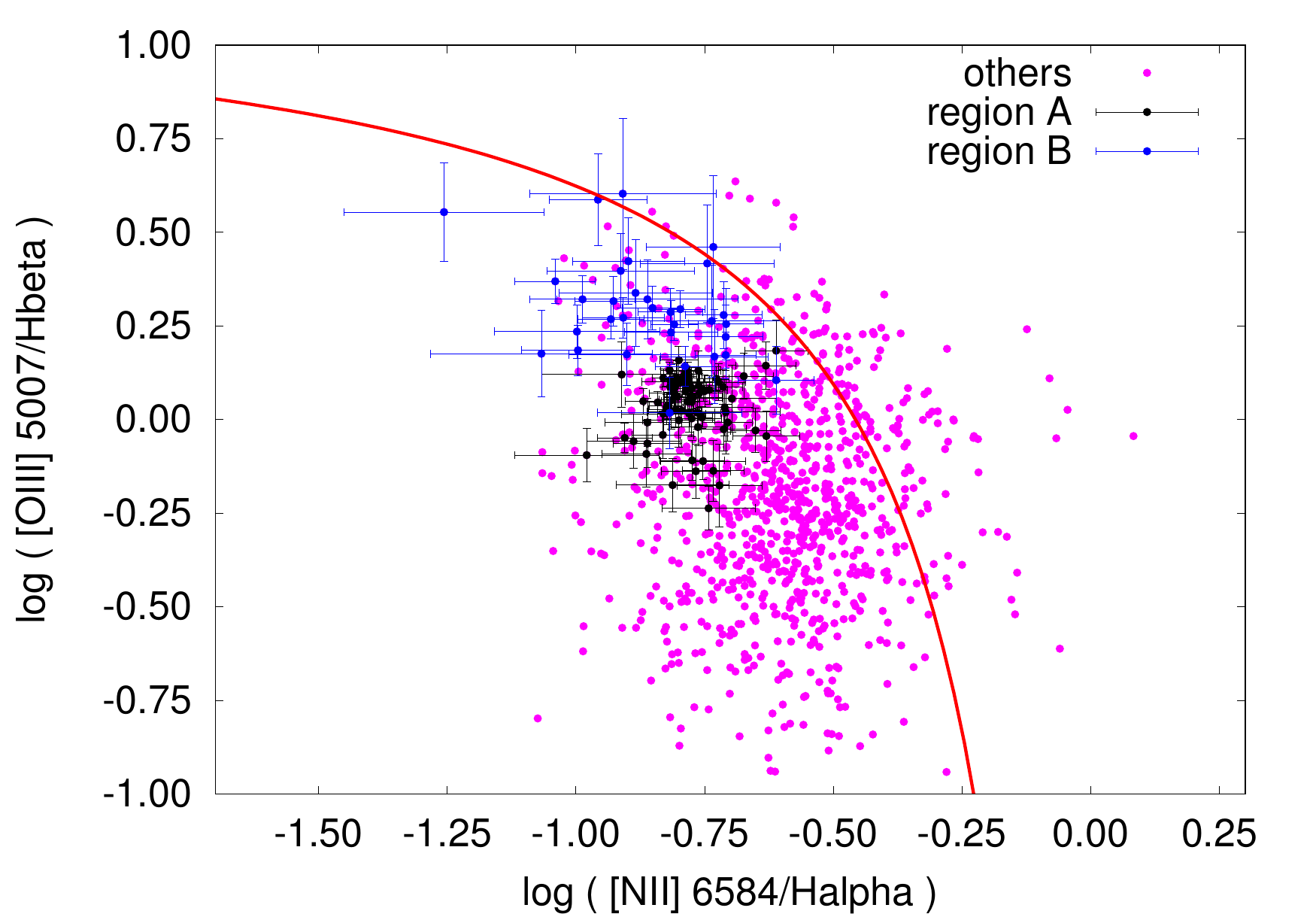}
\includegraphics[width=0.50\textwidth]{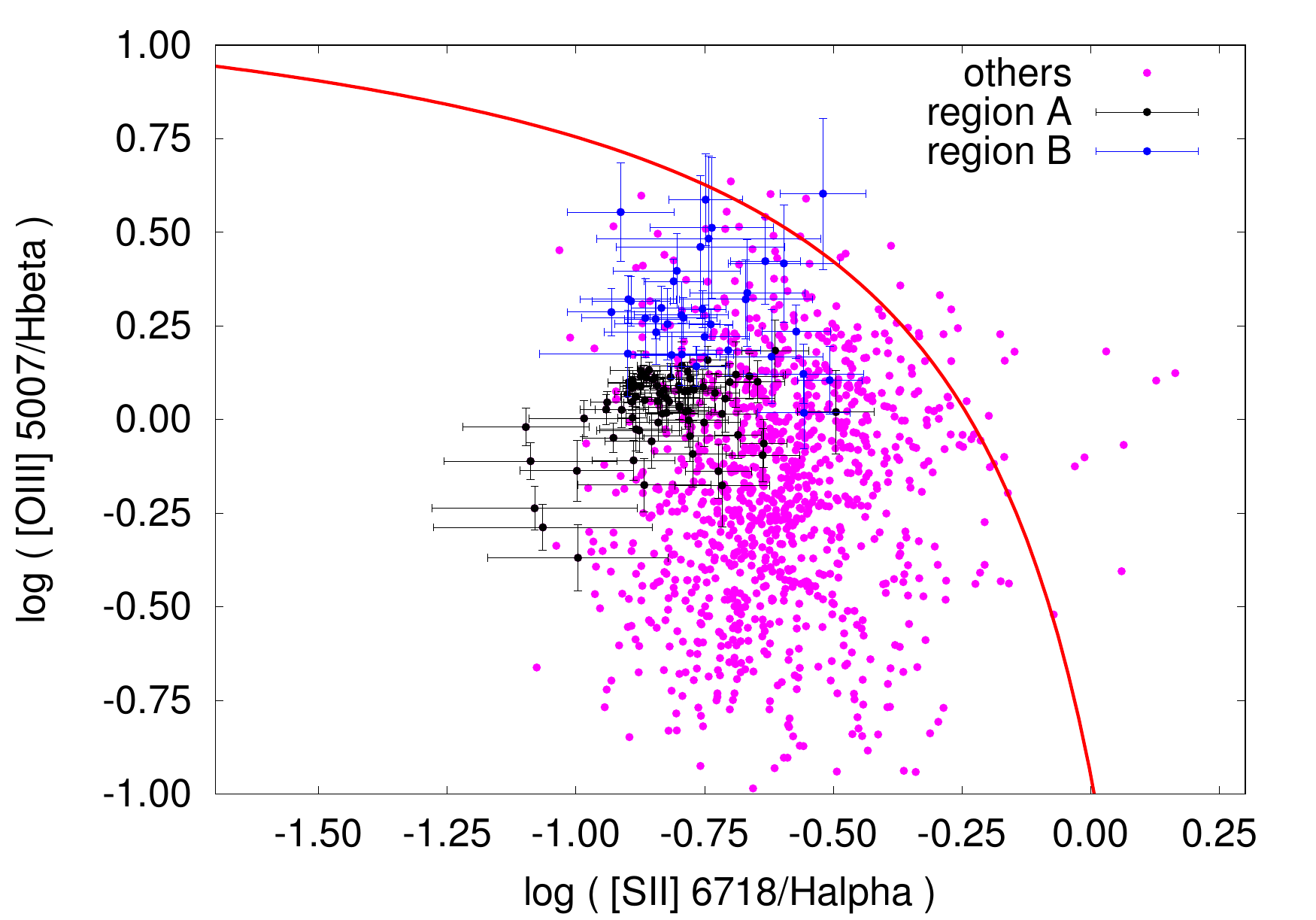}
\caption{
Emission-line Baldwin-Philips-Terlevich (BPT) diagnostic diagrams of
the host of GRB 080905A. Left: diagnostic diagram using
[N\,{\sc ii}].  Shown are the corresponding values for all individual
spaxels.  Black data points refer to spaxels that lie inside region A;
blue-colored data points refer to region B.  For the sake of clarity,
for these data points the corresponding $1\sigma$ error bars
are also shown.  The red line shows the pure star-formation demarcation
line for a $z$=0.1218 galaxy \citep[their
Equation 1]{Kewley2013ApJ...774L..10K}. Right: The same as left
but using [S\,{\sc ii}] (\citealt{Kewley2001ApJ...556..121K} and
\citealt{Kewley2006MNRAS.372..961K}; their Equations 6 and 2, respectively.).}
\label{fig:BPT}
\end{figure*}

Figure~\ref{fig:BPT} shows the loci in the BPT diagrams of the regions
with line emission in the GRB 080905A host galaxy. In both diagnostic
diagrams, the majority of spaxels fall inside the parameter space
characteristic of pure star formation. In the [N\,{\sc ii}] diagram,
star forming region A occupies a nearly circular area   centered at
the line ratios $(-0.780\pm0.006, +0.036\pm0.006)$;
in the [S\,{\sc ii}] diagram, the center of A has 
line ratios $(-0.827\pm0.008, +0.027\pm0.006)$. This
clustering is also seen in the [S\,{\sc ii}] diagram even though the
spread in data points is a little larger.  Not surprisingly, region A
lies in the pure star formation area of the diagrams. For the other
star-forming regions labeled in Fig.~\ref{fig:HST_080905} there are
significantly fewer data points and the clustering is less pronounced.
Therefore, as an example in (Fig.~\ref{fig:BPT}) we only show region
B.

\subsection{The Suspected Background Galaxy Close to the OT}

Using a circle with the diameter of the seeing disk (1 arcsec), at the
position of the suspected faint background galaxy mentioned by
\cite{Fong2013ApJ...776...18F} at R.A., decl. (J2000) = 19:10:41.743,
$-$18:52:48.15 ($\pm$0\farcs05; Sect.~\ref{background}), we detect
emission from the redshifted \Ha\ line at $z$=0.1218,
i.e., at the redshift of the large, barred spiral galaxy. There is also
evidence for two faint emission lines close to \Ha\, centered at 7340
and 7402~\AA~($\pm$0.5 \AA). Unfortunately, the nature of these two lines
could not be clarified. However, their origin in the suspected
background object is unlikely. The first emission-line feature is
also detected close to star-forming region A and the second feature
close to B. This is about 9\farcs2 and 13\farcs2, respectively, away from
the much less extended background object.

\section{Discussion \label{Discussion} }

Compared to the short-GRB host-galaxy ensemble listed in \citet[his
Table 2]{Berger2014ARAA}, an SFR of $\sim$1.6~\msunyr places
the host of GRB 080905A in the middle of a rather broad distribution
with SFR values between $<$0.1 and $>$10~$M_\odot$\,yr$^{-1}$.  Also
with respect to metallicity, the galaxy is a rather normal short-GRB
host.

Even though the global SFR of the GRB 080905A host is rather modest, the
MUSE data have revealed several star-forming complexes scattered
across the galaxy. Particularly striking is the giant star-forming
region A which is responsible for $\sim$20\% of the SFR  of the entire
galaxy and shines like a bright lighthouse in the \Ha\ line.
On the HST image, its
diameter is approximately twice the extension of the Tarantula nebula
(30 Doradus; diameter 370 pc) in the LMC
(\citealt{Crowther2019Galax...7...88C}), while its H$\alpha$
luminosity is nearly four times as high. 

The high-resolution HST image suggests that the angular size of this
banana-shaped region is only about 0\farcs3-0\farcs4 at its longest
(NW--SE) extension ($\sim$0.7-0.9 kpc) and consists of at least four
individual knots (Fig.~\ref{fig:HST_zoom}). The star formation rate
surface density $\Sigma_{\rm SFR}$ in this region is about
0.5-0.8~$M_\odot$\,yr$^{-1}$ kpc$^{-2}$. In the VLT/MUSE
data cube, this region cannot be spatially resolved.

To some degree, region A and its relation with its host resembles
the Wolf--Rayet star-forming region in the GRB 980425 host galaxy --
very bright in \Ha\ and a strong radiation field
(\citealt{Christensen2008A&A...490...45C,
Michalowski2014A&A...562A..70M, Michalowski2016A&A...595A..72M,
Kruehler2017}). GRB 980425 was a long burst, however; its origin
was the collapse of a massive star. In any case, very detailed
studies of long-GRB-hosts with VLT/MUSE (e.g., GRB
980425: \citealt{Kruehler2017}; GRB
100316D: \citealt{Izzo2017MNRAS.472.4480I}; GRB
111005A: \citealt{Tanga2018}) but also of the hosts of various
supernovae (e.g., \citealt{Chen2017ApJ...849L...4C,
Galbany2016MNRAS.455.4087G,Sun2021MNRAS.504.2253S}) have revealed a
plethora of information about the star-formation activity in dwarf and
spiral galaxies.

The center of region A lies about 10$''$ (22 kpc) in projection
away from the GRB explosion site.  Several other, but less luminous
star-forming regions are located at much closer distance to the
afterglow position. Could any of these star-forming regions be the
original birthplace of the short-GRB progenitor?  In order to tackle
this question, we made use of an available set of stellar population
synthesis calculations.

\subsection{The Stellar Population Synthesis Code \label{SPS}}

We employed the population synthesis code {\tt
StarTrack} \citep{Belczynski2002,Belczynski2008a}  to obtain
astrophysically motivated physical properties of NS--NS binaries.  We
adopted the rapid core-collapse supernova engine NS/BH mass
calculation \citep{Fryer2012} that allows for NS formation in a mass
range $1-2.5~M_\odot$.  Natal kicks NSs/BHs have received at formation are
taken from a one-dimensional Maxwellian distribution with $\sigma=265$~km
s$^{-1}$ (\citealt{Hobbs2005}) and were decreased inversely proportionally to
the amount of  fallback calculated for each supernova
event \citep{Fryer2012}. This procedure was applied
to NSs formed in core-collapse
supernovae. However, for NSs formed in electron-capture supernovae, we
did not apply a natal kick. Blaauw kicks (from a symmetric mass loss;
\citealt{Blaauw1961BAN....15..265B,
Repetto2012MNRAS.425.2799R}) were
applied to all NSs.  We assumed standard wind losses for massive stars:
O/B star (\citealt{Vink2001}) winds and LBV  winds \citep[specific
prescriptions for these winds are listed in Sect.~2.2
of][]{Belczynski2010b}.  We treated accretion onto a compact object
during Roche lobe overflow and from stellar winds  using the analytic
approximations presented in \cite{King2001,Mondal2020}. We adopted
a limited 5\% Bondi accretion rate onto BHs/NSs during common
envelope \citep{Ricker&Taam,MacLeod2017}.  The most updated
description of {\tt StarTrack} is given in
\cite{Belczynski2020b}. Here we use the input physics from model M30 of
that paper.  
Model 30 was used as it employs optimal input physics assumptions on
stellar and binary evolution as argued in \cite{Belczynski2020b}.

The numerical code delivers the NS masses, the time of their formation
(starting at ZAMS at $t$=0), the type of SN,
the time between the two SNe, the
velocity vector of the binary after the first ($t=t_1$) and after the
second SN ($t=t_2$), and the time from the second SN to the merger
($t=t_3$; Table~\ref{Tab:model}).  In other words, after the first SN
the system will begin moving in one direction for a time span
$t_2-t_1$. After the second supernova, the binary will start moving in
some other direction for a time span $t_3-t_2$.

\begin{figure}[t!]
\includegraphics[width=0.48\textwidth]{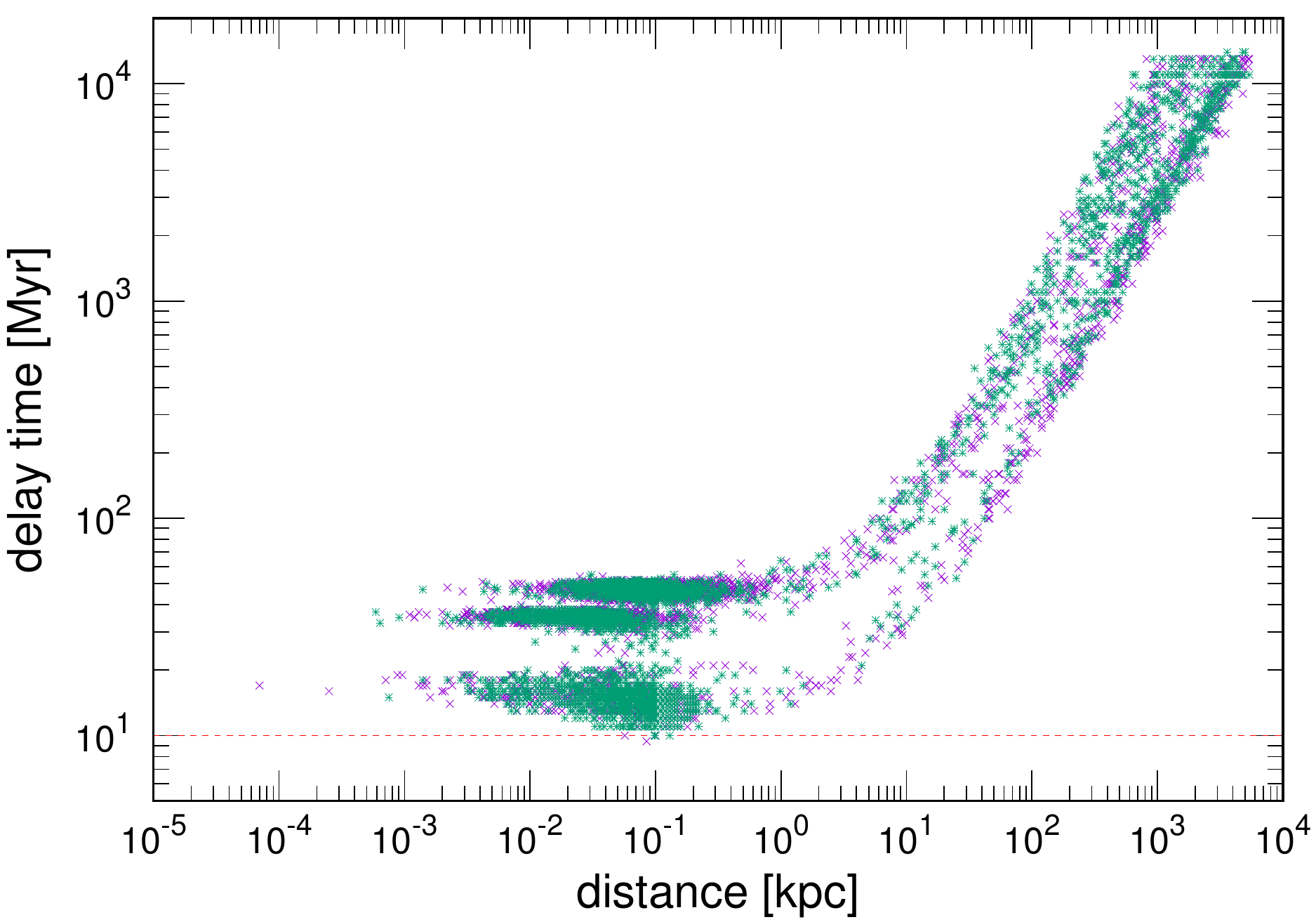}\\
\includegraphics[width=0.48\textwidth]{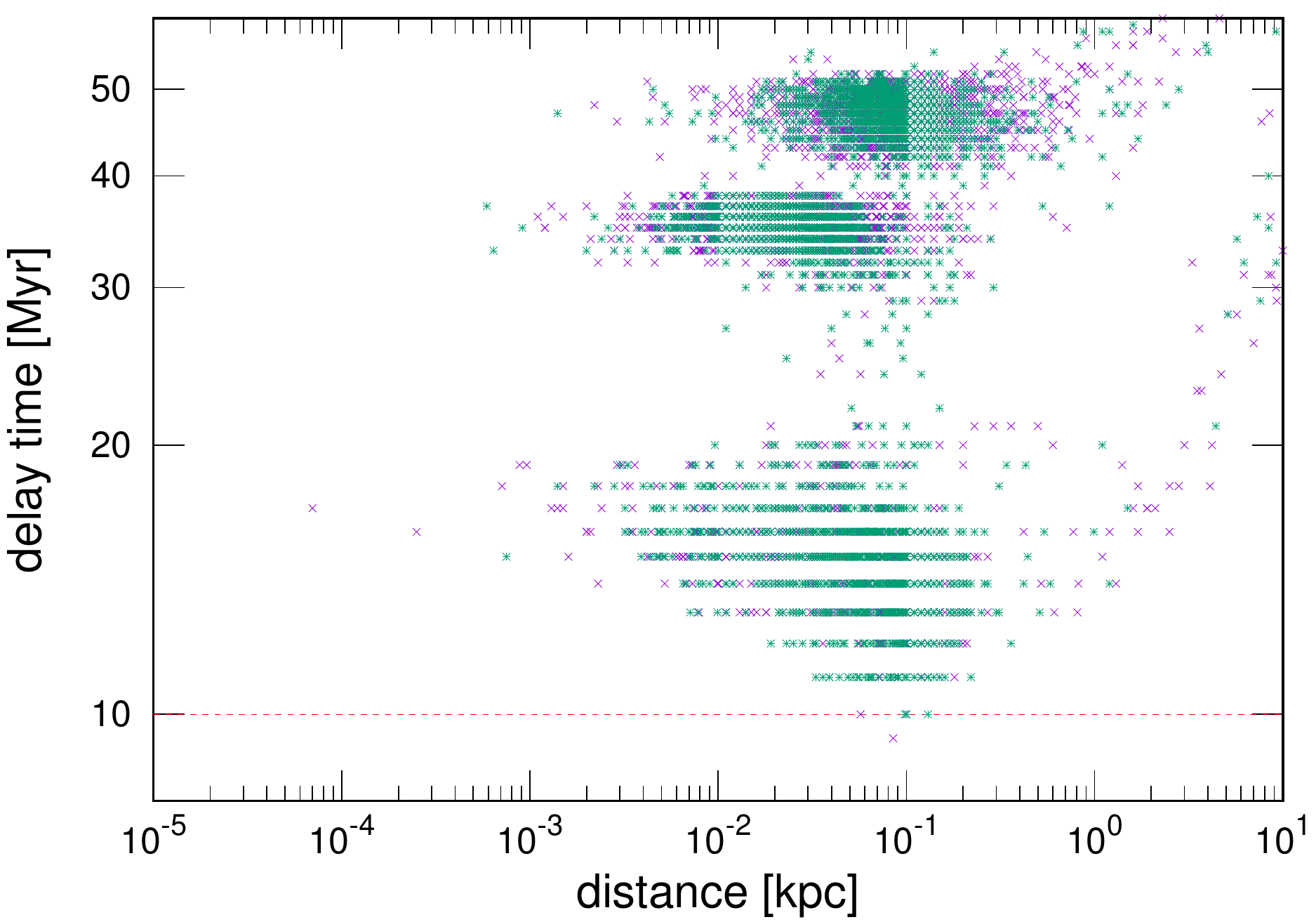}
\caption{
Top: delay time versus spatial distance from the stellar nursery
for the stellar population models considered here (green: model
M30, 3696 simulations; violet: model M35, 4633 simulations). All
but 1 NS--NS system are characterized by a delay time of more than 10
Myr (red line). Bottom: zoom-in into the upper figure to
better show the systems with the shortest delay times.
For the exceptional case with a delay time of 9.43 Myr, see
Sect.~\ref{appendix}. }
\label{fig:model1}
\end{figure}

\begin{deluxetable}{lll}
\tablecolumns{3} 
\tablecaption{The Stellar Population Synthesis Model.}
\tablewidth{0pt} 
\tablehead{
  \colhead{Time}    &
  \colhead{Distance}  &
  \colhead{Comment} } 
\startdata
$t$=0   &   $d$=0     &   ZAMS\\
$t=t_1$ &   $d=d_1$=0 &   1st SN\\
$t=t_2$ &   $d=d_2$   &   2nd SN\\
$t=t_3$ &   $d=d_3$   &   merger\\
\enddata
\tablecomments{The stellar population synthesis (SPS) model:
  the time difference $t_3 - t_2$ is the merger time
  $t_{\rm mr}$.  The delay time, however, is defined
  as the time from arrival at the ZAMS ($t$=0) to the merger: $t_{\rm
    delay}$ = max$(t_1, t_2) + t_{\rm mr}$.  $d$ is the spatial
  distance from the stellar nursery.}
\label{Tab:model}
\end{deluxetable}

Natal kicks and their mechanism are not fully understood or
observationally 
constrained, so, in order to fully assess this uncertainty, one way
could be to consider two extreme models: no natal kicks and maximum
(very hard to assess what is actual maximum) natal kicks (which are
estimated  at about $>$1000 km s$^{-1}$). This does not really need to
be shown, as no natal kicks  would result in travel distance close to
0 kpc (there would be a small systemic  velocity increase due to a
symmetric mass loss at NS formation), while for kicks as  high as 1000
km s$^{-1}$ there would be virtually no NS--NS progenitors
surviving. We  propose to do a different sort of possibly more realistic
exercise: if NSs in binaries  receive kicks, it is claimed that they 
receive possibly smaller kicks than  single pulsars
(\citealt{Willems2008AIPC..983..464W} and references
therein). Therefore, we also considered one more model,  exactly the
same as M30, but with 50\% natal kicks ($\sigma$ = 130 km s$^{-1}$;
model M35).  Figure~\ref{fig:model1} shows the arising
difference/uncertainty. It also shows that in all but one case, all
NS--NS mergers are characterized by a delay time larger than 10.0
Myr. The single exceptional case is an NS--NS system in model M30
with a delay time of 9.43 Myr (Sect.~\ref{appendix}).

\subsection{Linking the Model Results to the Host of the Short GRB 080905A}

Before applying the results of the numerical BNS population synthesis
model to the host case discussed here, we describe some specifics of
the assumed  star-formation history.  Following
\cite{Leitherer1999ApJS..123....3L}, we distinguish two scenarios.

\subsubsection{The Instantaneous Starburst Scenario}

Here we assume that all currently detectable H$\alpha$-bright
star-forming regions are the result of a single instantaneous
starburst. Within this scenario, all stars arrived at the ZAMS at
$t$=0. Then, after a while, SNe start to explode and compact stellar
binaries (NS--NS) begin to form and finally merge at a certain spatial
distance $d$ from the stellar birthplace (Table~\ref{Tab:model}).

Within the context of this model, the evolutionary state of a
star-forming region as a luminous H$\alpha$ source lasts less than
6-10 Myr (e.g., \citealt{Copetti1986,Tremblin2014A&A...568A...4T}).
As we have outlined in Section~\ref{EW}, using the the {\tt Starburst99}
database (\citealt{Leitherer1999ApJS..123....3L}), all
star-forming regions listed in Table~\ref{Tab:A1} have an age between
6 and 8 Myr.

The numerical modeling presented in Sect.~\ref{SPS} shows that very
special circumstances are required so that an NS--NS merger can occur
within a delay time of less than 10 Myr (see Sect.~\ref{appendix}).
We thus conclude that it is theoretically possible but statistically
not likely that an NS--NS binary merges already within 10 Myr
after the arrival of its two progenitor stars on the ZAMS. This
conclusion is insensitive to the adopted metallicity of the
star-forming region.

Adding to our model BH--NS mergers would not make any
difference to  these conclusions. (i) Their formation is less frequent
than  the formation of NS--NS systems (see Table 3 and 4 in
\citealt{Belczynski2020b} for model M30). (ii) Only a small
fraction of BH--NS mergers are predicted to  produce mass ejecta that
would power a GRB (see \citealt{Drozda2020arXiv200906655D}; BH--NS
mergers  with ejecta represent only $<$10\% of entire merging BH--NS
population). Very special conditions are required such as a high BH
spin and comparable masses of  BH and NS to produce mass ejecta in
BH--NS mergers. 

We can go one step further and assume that our age constraints that we
derived from {\tt Starburst99} are wrong by 50\%, i.e., we allow for stellar
ages up to 12 Myr. Then, in model M30, altogether 23 systems fulfill
this age criterion; in model M35 these are 69 binaries. However, if we
also  require that at the time of the merger the progenitor  system
has traveled at least 3 kpc from its birthplace, none of the binaries
satisfy this criterion.  All  NS--NS systems with short delay times
have travel distances between 0.06 and 0.18 kpc (model M30) and,
respectively, 0.02 and 0.36 kpc (model M35). This stems from the
fact that the shortest-lived NS--NS systems  cannot travel too far
even if they acquired high systemic speeds from natal kicks and/or SN
mass loss. In other words, taking into account the age and the
distance constraints, none of the simulated mergers allows for a link
of the merger progenitor to any of the presently active star-forming
regions listed in Table~\ref{Tab:A1}.  Within the context of the
instantaneous starburst scenario
(\citealt{Leitherer1999ApJS..123....3L}), none of these regions is a
likely birthplace of the short-GRB progenitor system.

We finally note that even within the context of the  instantaneous
starburst scenario, our MUSE observations do not exclude the potential
existence of star-forming regions older than 10 Myr. However, for such
ages, \Ha\ is no longer a good tracer of star-forming activity.  A
deep UV imaging of the suspected GRB 080905A host galaxy could reveal older
star-forming complexes, but no such data are at hand.

\subsubsection{The Continuous Star Formation Scenario}

In such a case, the interpretation of observational data becomes much
more complex. In this scenario, EW(\Ha) basically remains constant and at a
high level as long as the starburst holds on (e.g.,
\citealt{Kuncarayakti2013AJ....146...30K}, their figure 1).  For
example, if we increase the constraint on the age of the short-GRB
progenitor to 100 Myr, then 82\% of the modeled evolutionary tracks
predict an NS--NS merger within such a time span.  However, 
only a small fraction (2\%) of these NS--NS mergers with delay times
below 100 Myr have travel distances larger than 3 kpc.

\subsubsection{Binary Stellar Population Models}

{\tt Starburst99} is based on single-star evolutionary tracks.
Observations suggest, however, that most massive stars are born
in binaries (e.g., \citealt{Sana2012Sci...337..444S}).
Moreover, the progenitor of GRB 080905A might have been
a tight binary. This suggests that for EW(\Ha) age estimates,
binary stellar population models should be considered too.

\cite{Xiao2019MNRAS.482..384X} used the {\tt BPASS} (binary
population spectral and synthesis) models
(\citealt{Eldridge2017PASA...34...58E}) to investigate the
EW(\Ha)-age-metallicity relationship and compared their results with
single-star evolution models. They found that binary models can
predict a much larger age once EW(\Ha)$\lesssim$1000~\AA; and this holds for
all metallicities $Z\gtrsim0.001$. This age difference increases with
decreasing EW (see also \citealt{Lyman2018MNRAS.473.1359L}, their figure 4).
For $Z$=0.009--0.020 and EW(\Ha)=30--60~\AA, the predicted age range of
the {\tt BPASS} model is 20--40~Myr, compared to about 8--12~Myr for 
single-star models (\citealt{Kuncarayakti2016A&A...593A..78K},
their figure 1).

We find that in model M30 and in model M35, less than 1\% of all
NS--NS mergers have delay times smaller than 30 Myr
and travel distances larger than 3 kpc.
Similarly to the continuous star formation model,
the probability is low that the GRB progenitor fits into this
scenario even though it is clear that binary models can bring a new
flavor into the question we have touched here. Going further into
these models might be an option for further studies.
Discussing this issue in detail is, however, beyond the scope of this paper.

\section{Summary}

By combining archival HST data  (program ID 12502; PI: A.  Fruchter)
with VLT/MUSE observations, we have explored the star-formation
properties of the host of the short GRB 080905A. The MUSE data allowed
us to explore its internal extinction,  star forming activity,
metallicity, and its radial velocity pattern.

Using \Ha\ emission as a tracer for star formation, we found that the
host contains several luminous star-forming complexes scattered across
its spiral arms.  Star formation activity can be found throughout the
galaxy disk. The closest star-forming region to the GRB explosion site
is about 3 kpc (in projection) away.  For the entire galaxy, we measure
a \Ha\ SFR of about 1.6 \msunyr.

The largest and most luminous star-forming complex (labeled A;
Fig.~\ref{fig:HST_080905}) lies 22 kpc in projection away from the GRB
afterglow position. It shows a H$\alpha$ luminosity of about
$5\,\times\,10^{40}$ erg s$^{-1}$ and has a star formation rate
surface density $\Sigma_{\rm SFR}$ on the order of about
0.5--0.8~\msunyr per kpc$^2$. This is a high value and
comparable to what has been found in luminous infrared galaxies
(\citealt{Lopez2016A&A...590A..67P}).

For all star-forming regions, including region A, we measure an \Ha\
equivalent width that suggests individual ages of less than 10 Myr,
provided that all star-forming activity has its origin in single
instantaneous starbursts. Within this context, our stellar population
synthesis calculations show that none of these \Ha-bright star-forming regions
can be considered as a candidate stellar nursery that has formed the NS--NS
progenitor.

\begin{acknowledgements}

A.N.G. and S.K. acknowledge financial support by grants DFG Kl
766/16-3, DFG Kl 766/18-1, and DFG RA 2484/1-3.  P.S.
acknowledges support through the Sofja Kovalevskaja Award
from the Alexander von Humboldt Foundation of Germany.
K.B. acknowledges support from the Polish National Science Center grant
Maestro (2018/30/A/ST9/00050).
M.J.M.~acknowledges the support of the National Science Centre,
Poland through the SONATA BIS grant 2018/30/E/ST9/00208.
The authors thank Thomas Kr\"uhler for providing several numerical
tools for VLT/MUSE data reduction and for taking care of the VLT/MUSE
observations and a first data reduction.
%
Based on observations made with the NASA/ESA Hubble Space Telescope
and obtained from the Hubble Legacy Archive, which is a collaboration
between the Space Telescope Science Institute (STScI/NASA), the Space
Telescope European Coordinating Facility (ST-ECF/ESA), and the Canadian
Astronomy Data Centre (CADC/NRC/CSA).
The Australia Telescope is funded by the Commonwealth of Australia for
operation as a National Facility managed by CSIRO.  A.N.G. and S.K. thank
Catarina Ubach \& Sarah Maddison, Swinburne University, Mark Wieringa
and Ivy Wong, CSIRO Sydney, Jamie Stevens, CSIRO Narrabri,
and Martin Bell, now University of Technology, Sydney,
for helpful discussions and observing guidance.
This work made use of data supplied by the UK Swift Science Data
Centre at the University of Leicester.
We thank the anonymous referee for a very careful reading of the manuscript
and for many very valuable suggestions that helped to improve
the paper.

\end{acknowledgements}

\software{MIRIAD (\citealt{Sault1995}), DS9 (ver8.1; \citealt{Joye2003ASPC..295..489J}),
Starlight (\citealt{Cid2005, Cid2009}), StarTrack (\citealt{Belczynski2002,Belczynski2008a})}.


\begin{appendix}
  
\onecolumngrid

\newpage

\section{Additional Tables} 

Table 4 provides for all regions indicated in Fig.~\ref{fig:HST_080905}
the measured luminosities in the five emission lines.

\begin{deluxetable*}{rrrrrr}[h!]
\tablecolumns{9} 
\tablewidth{-12.0pt} 
\tablecaption{Measured Emission-line Luminosities (in Units of 10$^{39}$ erg s$^{-1}$)
  for the Regions Indicated in Figure~\ref{fig:HST_080905}.}
\tablehead{
  \colhead{Region  }  &
  \colhead{$L$(\Ha)}  &
  \colhead{$L$(\Hb) }  &
  \colhead{$L$(\OIII 5007)} &
  \colhead{$L$(\NII 6584)} &
  \colhead{$L$(\SII 6718)} }
\startdata
   A       &  45.20 $\pm$ 0.22 & 14.40 $\pm$ 0.23 &  12.60 $\pm$ 0.22 & 5.77 $\pm$ 0.21 & 5.14 $\pm$ 0.21  \\
   B       &   7.86 $\pm$ 0.15 &  2.33 $\pm$ 0.15 &   3.25 $\pm$ 0.15 & 0.79 $\pm$ 0.14 & 1.04 $\pm$ 0.14  \\
   C       &   4.81 $\pm$ 0.12 &  1.05 $\pm$ 0.12 &   0.15 $\pm$ 0.11 & 1.45 $\pm$ 0.11 & 1.06 $\pm$ 0.11  \\
   D       &  14.40 $\pm$ 0.21 &  4.04 $\pm$ 0.21 &   1.31 $\pm$ 0.20 & 3.12 $\pm$ 0.19 & 2.54 $\pm$ 0.20  \\
   X       &   5.26 $\pm$ 0.25 &  1.44 $\pm$ 0.25 &   1.70 $\pm$ 0.25 & 0.52 $\pm$ 0.24 & 0.80 $\pm$ 0.24  \\[1mm]
   1       &   1.63 $\pm$ 0.10 &  0.57 $\pm$ 0.10 &   0.66 $\pm$ 0.10 & 0.12 $\pm$ 0.10 & 0.32 $\pm$ 0.10  \\
   2       &   4.27 $\pm$ 0.11 &  1.62 $\pm$ 0.11 &   0.70 $\pm$ 0.11 & 0.80 $\pm$ 0.10 & 0.71 $\pm$ 0.10  \\
   3       &   3.09 $\pm$ 0.11 &  0.97 $\pm$ 0.11 &   0.51 $\pm$ 0.10 & 0.53 $\pm$ 0.10 & 0.54 $\pm$ 0.10  \\
   4       &   6.92 $\pm$ 0.11 &  2.38 $\pm$ 0.11 &   0.74 $\pm$ 0.10 & 1.48 $\pm$ 0.10 & 1.10 $\pm$ 0.10  \\
   5       &   2.75 $\pm$ 0.11 &  0.87 $\pm$ 0.11 &   0.85 $\pm$ 0.10 & 0.36 $\pm$ 0.10 & 0.52 $\pm$ 0.10  \\
   6       &   6.15 $\pm$ 0.11 &  3.21 $\pm$ 0.11 &   0.88 $\pm$ 0.11 & 1.26 $\pm$ 0.10 & 0.85 $\pm$ 0.10  \\
   7       &   1.77 $\pm$ 0.11 &  0.61 $\pm$ 0.11 &   1.05 $\pm$ 0.10 & 0.10 $\pm$ 0.10 & 0.19 $\pm$ 0.10  \\
   8       &   2.54 $\pm$ 0.11 &  0.75 $\pm$ 0.11 &   0.85 $\pm$ 0.10 & 0.25 $\pm$ 0.10 & 0.32 $\pm$ 0.10  \\
   9       &   2.14 $\pm$ 0.11 &  0.46 $\pm$ 0.11 &   0.32 $\pm$ 0.10 & 0.43 $\pm$ 0.10 & 0.38 $\pm$ 0.10  \\
   10      &   2.42 $\pm$ 0.11 &  0.94 $\pm$ 0.11 &   0.79 $\pm$ 0.10 & 0.30 $\pm$ 0.10 & 0.36 $\pm$ 0.10  \\[1mm]
\hline                                                                                                       
  galaxy   & 264.00 $\pm$ 1.38 & 88.00 $\pm$ 1.37 &  59.60 $\pm$ 1.33 &40.90 $\pm$ 1.32 &42.10 $\pm$ 1.32  \\
\enddata
\tablecomments{
The luminosities are based on all spaxels which fulfill the criterion
S/N$\geq$2 in the corresponding line. The data for the entire galaxy
exclude the flux coming from masked regions that cover the three bright
Galactic foreground stars (Fig.~\ref{fig:HST_080905}). With the
exception of \Hb, for these three regions we used a mask with a radius
of 0\farcs7. In the case of \Hb, for the two stars next to region A
we had to increase the radius of the mask to 0\farcs9; for the third
star next to regions \#2 and \#4 we still used 0\farcs7, however. The error
includes the r.m.s. of the corresponding luminosity, measured in an
area with the diameter of the seeing disk (1 arcsec) in various regions
around the galaxy and added in quadrature to the measurement
error. For all lines this $1\sigma_{\rm rms}$ error was about $10^{37}$
erg s$^{-1}$.}
\label{Tab:Appendix}
\end{deluxetable*}

\newpage
.
\newpage

\twocolumngrid

\section{An NS--NS Model With a Delay Time of Less than 10 Myr \label{appendix} }

In the SPS model outlined in Sect.~\ref{SPS},
the minimum timescale is set by the time the two massive stars need to evolve
from the ZAMS to the formation of two NSs. The time needed for the two NSs to
merge (due to gravitational wave emission) may be very short for very
close (common  envelope) and highly eccentric orbits (natal
kick). The evolutionary time is set by the stellar mass. In our models single
stars with a mass below $\sim20$\msun form NSs while heavier stars form
BHs. The evolution of a 20\msun star takes $\sim10.5$~Myr. Binary evolution
may in exceptional cases act in such way as to produce NS--NS mergers
at timescales 
below 10~Myr. In our particular case of an NS--NS merger with a delay  time
of 9.4~Myr, this is exactly what happens. 

The binary system starts with two massive stars above the NS formation mass
limit: 22.34\msun and 22.33\msun at ZAMS. Note that these stars have
virtually the same mass which is unusual as our initial binary
population has a uniformly distributed  mass ratio. The more massive star
(primary) begins Roche lobe overflow (RLOF) on a thermal  timescale
right after leaving the main sequence and a large fraction of its envelope
is lost from the binary. Soon after, the less massive star (secondary)
also expands  beyond its Roche lobe and initiates a common envelope (CE)
that removes its H-rich  envelope and whatever is left from the H-rich
envelope of the primary star. Then at a time $t=9.1$ Myr, the secondary
explodes in a core-collapse SN forming a more massive NS in the system:
1.84\msun. Note two things: (i) the secondary explodes  first
due to a mass ratio reversal caused by mass accretion from the primary
during RLOF,  and (ii) NS formation happens below $t$=10 Myr
due to the fact that the main-sequence evolution timescale was well below
that of the most massive single star that can form an NS.\footnote{An NS
is formed anyway despite the fact that its
initial stellar mass was well above the NS
formation mass due to the loss of the star's entire H-rich envelope right
after the main sequence.} Shortly after, the primary expands again and as a
helium giant initiates yet another CE phase, it loses its entire
He-rich  envelope. At this point, it lost enough mass so that it also
forms an NS with a mass of 1.44\msun and the second supernova happens at
$t=9.4$ Myr. At this point, the system is very compact (after two CE
phases) so that it may easily survive even a large natal kick that will
induce significant eccentricity to the NS--NS binary without disrupting the
system. This significantly reduces the merger time.  The final NS--NS
semi-major axis is very small ($a=0.3$\rsun) and eccentric ($e=0.6$)
and that translates to a merger time of only 0.03 Myr. Therefore, it
takes only 9.43 Myr from ZAMS to the NS--NS merger and the GRB. The
combination of very high stellar masses and the fact of both masses
being virtually equal  sends the system on the very unique
evolutionary trajectory generating a very  short delay time.

\section{ATCA Constraints on GRB Late-time Emission Components}

The ATCA radio observations constrain the luminosity of any long-lived
radio transient related to the burst.  This refers to the luminosity
of the radio afterglow (e.g., \citealt{Chandra2012})  and to a
potential late-time kilonova radio flare
(\citealt{Nakar2011,Metzger2014MNRAS.437.1821M,Margalit2015,
Fong2016ApJ,Horesh2016,Radice:2018pdn}).

Following the procedure outlined in \cite{Klose2019}, and assuming an
isotropically radiating source, the corresponding upper limits are
listed in Table~\ref{Tab:KN}.  Here, for the radio afterglow, we
considered a spectral slope ($F_\nu \sim \nu^{-\beta}$) of
$\beta=-1/3$ or 0.7 and for a kilonova radio flare, we set $\beta=0.7$.

Since our ATCA observations resulted in non-detections, we provide the
corresponding upper limits on the luminosity based on multiples of the
noise level  on the radio image. In the Table we provide the measured
$1\sigma$ r.m.s. in an area of $60''\times60''$ centered at the target
position for a robust parameter of 0.5.  Following \cite{Klose2019}
and \cite{Nicuesa2020}, we then calculated the corresponding radio
luminosity upper limits of a point source by using five times this
r.m.s. value. For the sake of clarity we note that using our data we
could also claim $2\sigma$ luminosity upper limits. However, given the
fact that these are interferometric data and the noise is not really
Gaussian, we follow others (e.g.,
\citealt{Wadadekar1999AJ....118.1435W,Tasse2006A&A...456..791T,
Stroe2012A&A...546A.116S}) and provide $5\sigma$ sensitivities.  We
refer to \cite{Murphy2017ApJ...839...35M} for a more stringent
discussion on this important issue.

\begin{deluxetable*}{rccr cccc}
\tablecolumns{8} 
\tablewidth{5.0pt} 
\caption{$5\sigma$ Upper Limits on Late-time Radiation Components.}
\tablehead{
\colhead{Run } &
\colhead{$dt_{\rm obs}$ } &
\colhead{$dt_{\rm host}$ } &
\colhead{$1\sigma_{\rm rms}$ } &
\colhead{$L_{\nu, 1}$ } &
\colhead{$L_{\nu, 2}$ } &
\colhead{$\nu L_{\nu, 1}$} &
\colhead{$\nu L_{\nu, 2}$}\\
\colhead{(1)} &
\colhead{(2)} &
\colhead{(3)} &
\colhead{(4)} &
\colhead{(5)} &
\colhead{(6)} &
\colhead{(7)} &
\colhead{(8)} }
\startdata
Run 1, 5.5 GHz & 4.87 & 4.34 & 12.6 &   2.2 &   2.5 &   1.2 &   1.4\\
       9.0 GHz &      &      & 15.4 &   2.7 &   3.0 &   2.4 &   2.7\\
Run 2, 5.5 GHz & 7.13 & 6.36 &  6.7 &   1.2 &   1.3 &   0.6 &   0.7\\
       9.0 GHz &      &      &  8.6 &   1.5 &   1.7 &   1.4 &   1.5\\
\enddata
\tablecomments{
Columns \#2 and \#3 give the time after the burst in the observer and
in the host frame, respectively. Column \#4 contains the   1$\sigma$
r.m.s. on the radio image. Columns \#5 to \#8 provide the corresponding
5$\sigma$ upper limits (in units of $\mu$Jy beam$^{-1}$) for the
specific luminosity $L_\nu$ (in units of $10^{28}$ erg s$^{-1}$
Hz$^{-1}$) as well as for $\nu L_\nu$ (in units of $10^{38}$ erg
s$^{-1}$), assuming isotropic emission. $L_{\nu, 1}$ assumes a
spectral slope $\beta =-1/3$, $L_{\nu, 2}$ assumes $\beta = 0.7$. }
\label{Tab:KN}
\end{deluxetable*}

The non-detection of the radio afterglow is not surprising given the
statistics of the observed luminosities of radio afterglows from
short GRBs (\citealt{Chandra2012, Fong2015ApJ...815..102F}).  The
non-detection of a late-time radio flare is in the line with other
unsuccessful searches for such signatures
(\citealt{Metzger2014MNRAS.437.1821M,
  Fong2016ApJ,Horesh2016,
  Klose2019,
  Ricci2021MNRAS.500.1708R,
  Schroeder2020ApJ...902...82S}).
For a detailed theoretical discussion of these non-detections
we refer to recent papers by \cite{Margalit2020MNRAS.495.4981M} and
\cite{Liu2020ApJ...890..102L}.

\end{appendix}


\bibliographystyle{aa_url}
\bibliography{mypaper}

\end{document}